\documentclass[fleqn,usenatbib]{mnras}

\usepackage{newtxtext,newtxmath}

\usepackage[T1]{fontenc}

\usepackage{ulem}

\DeclareRobustCommand{\VAN}[3]{#2}
\let\VANthebibliography\thebibliography
\def\thebibliography{\DeclareRobustCommand{\VAN}[3]{##3}\VANthebibliography}

\usepackage{graphicx}	
\usepackage{amsmath}	

\usepackage[dvipsnames]{xcolor}

\usepackage{placeins}

\newcommand{\metallicity}{$\log Z/Z_{\sun}$}
\newcommand{\NHI}{N_{\ion{H}{I}}}
\def\kms{\hbox{km~s$^{-1}$}}

\title[``Observing'' Synthetic Circumgalactic Absorption Spectra]{The Halo21 Absorption Modeling Challenge:\\Lessons from ``Observing'' Synthetic Circumgalactic Absorption Spectra}

\author[Hafen, Sameer, et al.]{
\parbox{\textwidth}{
Zachary Hafen$^{1}$,
Sameer$^{2,3}$,
Cameron Hummels$^4$,
Jane Charlton$^2$,
Nir Mandelker$^{5, 6}$,
Nastasha Wijers$^{7, 8}$,
James Bullock$^{1}$,
Yakov Faerman$^{9}$,
Nicolas Lehner$^{3}$,
Jonathan Stern$^{10}$
} \vspace{0.4cm}\\
\parbox{\textwidth}{
$^1$ Department of Physics and Astronomy, University of California Irvine, CA 92697, USA \\
$^{2}$ Department of Astronomy \& Astrophysics, 525 Davey Lab, The Pennsylvania State University, University Park, PA 16802, USA \\
$^{3}$ Department of Physics \& Astronomy, Nieuwland Science Hall, The University of Notre Dame, Notre Dame, IN 46556, USA \\
$^4$ TAPIR, California Institute of Technology, Pasadena, CA 91125, USA \\
$^5$ Racah Institute of Physics, The Hebrew University of Jerusalem,
Jerusalem 91904, Israel \\
$^6$ Kavli Institute for Theoretical Physics, Kohn Hall, Santa Barbara, CA 93106, USA\\
$^7$ Leiden Observatory, Leiden University, PO Box 9513, NL-2300 RA Leiden, The Netherlands \\
$^8$ Center for Interdisciplinary Exploration and Research in Astrophysics (CIERA), Northwestern University, 1800 Sherman Ave, Evanston, IL 60201, USA \\
$^9$ Astronomy Department, University of Washington, Seattle, WA 98195, USA \\
$^{10}$ School of Physics \& Astronomy, Tel Aviv University, Tel Aviv 69978, Israel
}
}

\date{Accepted XXX. Received YYY; in original form ZZZ}

\pubyear{2023}

\begin{document}
\label{firstpage}
\pagerange{\pageref{firstpage}--\pageref{lastpage}}
\maketitle

\begin{abstract}
In \textit{the Halo21 absorption modeling challenge} we generated synthetic absorption spectra of the circumgalactic medium (CGM),
and attempted to estimate the metallicity, temperature, and density ($Z$, $T$, and $n_{\rm H}$) of the underlying gas using observational methods.
We iteratively generated and analyzed three increasingly-complex data samples:
ion column densities of isolated uniform clouds,
mock spectra of 1--3 uniform clouds,
and mock spectra of high-resolution turbulent mixing zones.
We found that the observational estimates were accurate for both uniform cloud samples, with $Z$, $T$, and $n_{\rm H}$ retrieved within $0.1$ dex of the source value for $\gtrsim 90\%$ of absorption systems.
In the turbulent-mixing scenario, the mass, temperature, and metallicity of the strongest absorption components were also retrieved with high accuracy.
However, the underlying properties of the subdominant components were poorly constrained because the corresponding simulated gas contributed only weakly to the \ion{H}{I} absorption profiles.
On the other hand, including
additional components beyond the dominant ones did improve the fit, consistent with the true existence of complex cloud structures in the source data. 
\end{abstract}

\begin{keywords}
galaxies: absorption lines -- galaxies: haloes -- methods: data analysis
\end{keywords}



\section{Introduction}

With the understanding that galaxies and their environments are part of an interconnected galactic ecosystem, one of the primary goals of galaxy-scale astrophysics for the next decade is to identify the processes that drive galaxy growth~\citep{Decadal2020}.
Central to this effort are absorption system observations~\citep[e.g.][]{bahcall1993Hubble, lanzetta1995Gaseous, lauroesch1996QSO, Charlton1996,churchill1996Spatial,Prochaska1997,Rauch1997,Tumlinson2013,Werk2014,Prochaska2017,Kacprzak2019,Lehner2020}, one of the only constraints on the gaseous atmospheres of galaxies (the circumgalactic medium; CGM) as well as the intergalactic medium (IGM) between galaxies.
Absorption system observations typically consist of $\sim 1$ absorption spectrum per halo, each from a beam of background light (typically a quasar) that is partially absorbed by intervening gas en route to the observer.
These observations provide essential data, but are also notoriously difficult to interpret.

The challenge in interpreting absorption spectra arises in two ways.
The first challenge is that it is highly non-trivial to extract the properties of the gas responsible for a given absorption spectrum.
In observations the properties of the gas are derived based on spectra of individual ions (e.g. \ion{H}{I}, \ion{Mg}{II}, and \ion{O}{VI}), and their interpretation is hampered by line saturation, large and uncertain ionization corrections~\citep[e.g.][]{schaye2006Importance, acharya2021How}, and more.
One of the largest uncertainties is the structure of the absorbing gas---the simplest assumption is that the absorbing gas is a single cloud with uniform temperature, density, and metallicity, but many absorption spectra are best fit by assuming multiple clouds spanning a range of properties~\citep[e.g.][]{boksenberg1979Multiple, muzahid2015Extreme, liang2017BayesVP, Lehner2019,Wotta2019, haislmaier2021COS, sameer2021Cloudbycloud, zahedy2021.CUBS.III.zle1.LLSs, marra2021.cosmo.sims.test.observational.modeling, narayanan2021.a.multiphase.pLLS, nielsen2022Complex}, with possible overlapping spectra from physically-separated clouds~\citep[e.g.][]{marra2022Examining}.
High-resolution simulations and ionization models predict a distribution of many small clouds instead of a few larger clouds~\citep[e.g.][]{fielding2020Multiphase, lehner2019COS},
consistent with nearby ($\lesssim 100$~pc to the sun) interstellar absorption systems best fit by multiple components~\citep[e.g.][]{welsh2010HighResolution}.

The second challenge is that the CGM is expected to be a highly complex, multiphase system that pushes the limits of available information.
The information obtainable from an absorption spectrum is typically restricted to the temperature, density, and metallicity as a function of line-of-sight velocity along $\ll 100$ spectral skewers through the CGM of a given galaxy.
If the CGM is described by a few discrete clouds (as proposed by earlier models for absorption system distributions, e.g.~\citealt{srianand1994Halo, das2001Unified, maller2003Damped}) a small number of sightlines per halo might be sufficient to constrain models of the CGM.
Instead, observations suggest metallicities, densities, and temperatures vary over orders of magnitude across the CGM of different galaxies~\citep[e.g.][]{Lehner2019, Lehner2022} as well as within the CGM of a single galaxy~\citep[e.g.][]{lehner2020Project}.
This is consistent with analytic theory and high-resolution simulations that predict fragmentation and mixing of circumgalactic and intergalactic clouds~\citep[e.g.][]{maller2004Multiphase, mccourt2018Characteristic, hummels2019Impact, vandevoort2019Cosmological, peeples2019Figuring, mandelker2019Shattering, Mandelker.etal.2021}.
A complex multiphase CGM is also seen in cosmological simulations, which predict wind from the central galaxy, wind and stripping from satellite galaxies, and pristine accretion all give rise to absorbing gas, and in fact multiple origins likely overlap along a given sightline~\citep[e.g.][]{hafen2019Origins, hafen2020Fates, saeedzadeh2023Cool}.
As such, even given perfect information from each observed absorption line spectrum it may be challenging to use that information to gain a clear picture of cosmic ecosystems.

Despite the above concerns, there is reason to be optimistic,
as interpretations of absorption spectra continue to improve~\citep[e.g.][]{churchill2015Direct, sameer2021Cloudbycloud}, often with insight from synthetic spectra~\citep[e.g.][]{hummels2013Constraints, liang2018Observing}.
This paper marks the next step in improving interpretations of absorption spectra---the Halo21 Absorption Modeling Challenge, a mock-data challenge~\citep[e.g.][]{regimbau2012Mock, meacher2015Mock, hazboun2019Second} that combines the expertise of observers and theorists to test the limits of the current methodology for interpreting absorption spectra.

Attendees of the Halo21 KITP virtual conference\footnote{Halo21 (Hummels et al., in prep) was a ten-week open-attendance online conference held in early 2021 with the support of the Kavli Institute for Theoretical Physics at Santa Barbara.}
showed significant support in addressing the above issues, and subsequently we organized the Halo21 Absorption Modeling Challenge.
We invited the conference's 279 attending scientists to participate in one of two groups:
theorists and observers.
The premise of the Halo21 Absorption Modeling Challenge was for the theorists to generate mock data (ion column densities and spectra), and for the observers to estimate the source metallicity, density, and temperature of the provided mock data.
By doing so we aimed to improve our understanding of the uncertainties and systematics involved in absorption-system parameter estimation.
This allows for a validation of analysis techniques in isolation from real spectra,
which may be affected by higher-amplitude noise, fixed-pattern noise, and other contaminations.
The modeling challenge consisted of the following procedure:
(1) theorists decide on and generate the synthetic spectra for the observers to analyze; (2) observers make their best effort at estimating the parameters of the underlying gas which produced the spectra; and (3) finally we reveal the underlying gas properties, compare the results, and revise the parameter estimation to improve agreement.
In \S\ref{s: data generation} we describe the data samples, in order of increasing complexity, and the process for generating them.
In \S\ref{s:  parameter estimation} we describe the absorption system modeling process.
In \S\ref{s: results} we compare the modeled properties to the actual properties, and discuss the results in \S\ref{s: discussion}.
We conclude in \S\ref{s: conclusions}.
Throughout the text we associate the color black with the source data and
blue with the parameter estimates.
Throughout the paper $\log$ refers to $\log_{10}$.

\section{Data generation and parameter estimation}
\label{s: data generation}

\begin{figure*}
    \centering
    \includegraphics[height=0.9\textheight]{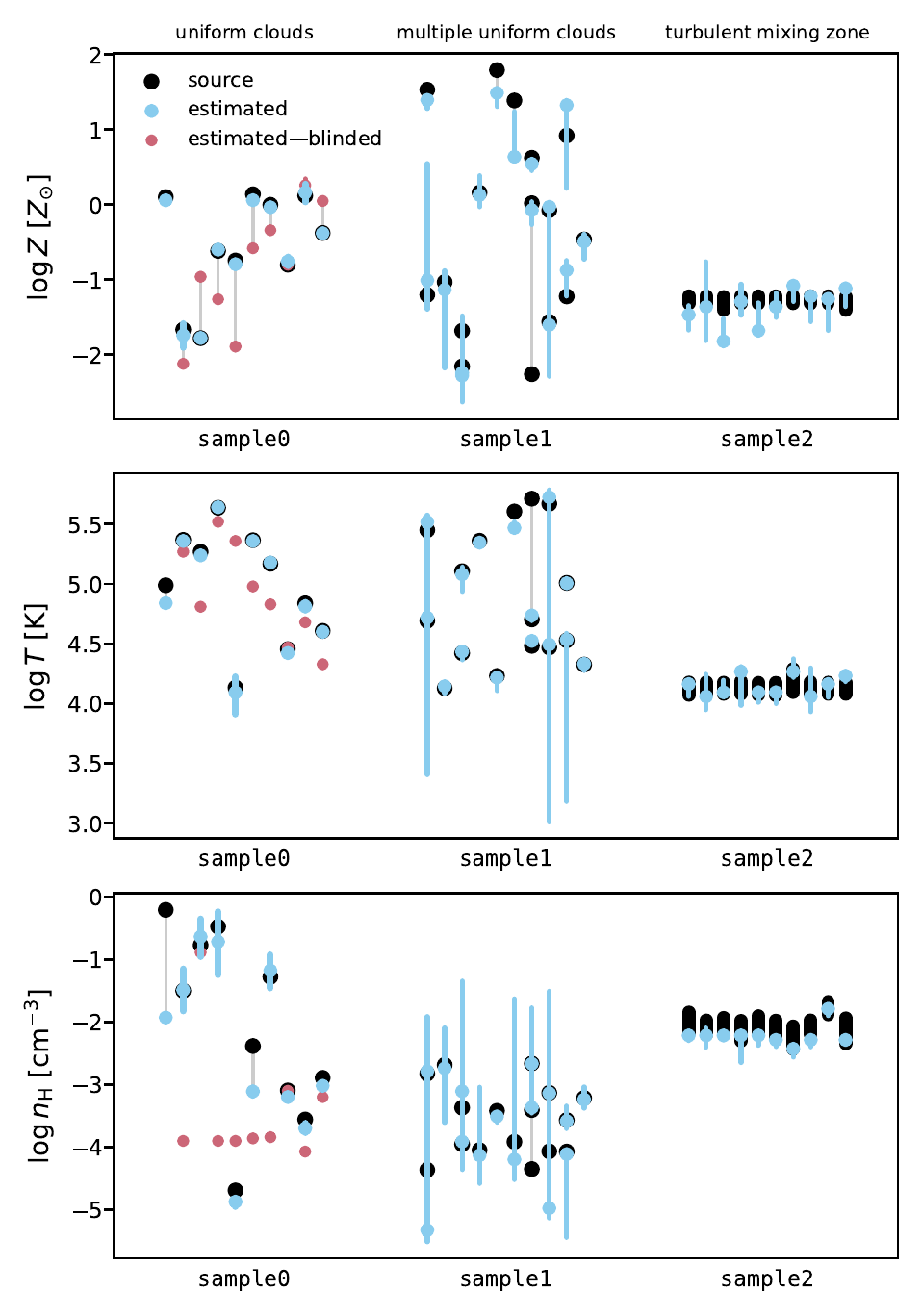}
    \caption{
    The source properties (black) used to produce the synthetic absorption systems
    compared to the best estimates from parameter estimation (blue) across all three of our samples.
    The lines span the 16th to 84th percentiles of the parameter-estimation posteriors or source data when available.
    The samples range from uniform clouds to a turbulent mixing zone.
    For the turbulent mixing zone the thick black line spans the parameter range responsible for $68\%$ of the \ion{H}{I}-absorption.
    The parameter-estimation best estimates and the source data agree to $\lesssim 0.1$ dex in ( $Z$, $T$, $n_{\rm H}$) for $\gtrsim 90\%$ of the systems,
    excluding the blinded \texttt{sample0} estimates (red; \S\ref{s: results -- sample0}).
    The grey lines connect the source clouds and the corresponding estimates.
    }
    \label{f: summary}
\end{figure*}

Participating theorists came from a wide variety of CGM-related backgrounds, from cosmological and idealized simulations to analytic models.
Drawing on this expertise, theorists produced three mock data samples of increasing sophistication for observers to model:
a sample of ion column densities for uniform clouds,
a sample of multi-phase spectra intersecting one to three clouds per absorption system,
and a sample of spectra drawn from a high-resolution simulation of a $T \sim 10^4$ K filament embedded in a $T \sim 10^6$ K halo.
The metallicity, density, and temperature of all samples are shown in Figure~\ref{f: summary}.

During the production of all three samples, we used the synthetic spectrum generator \textsc{trident}~\citep{hummels2017Trident}.  \textsc{trident} is a Python package for creating synthetic absorption-line spectra from the distribution of gas in theoretical models and hydrodynamic simulations. It post-processes these data to approximate the abundance of different ions due to collisional and photo-ionization processes.  The user selects an arbitrary sightline through the gas distribution to represent the path of a photon from a background quasar on its way to the synthetic telescope.  Trident generates a corresponding synthetic spectrum, depositing Voigt profiles (VP) for each desired absorption line, as it steps through intervening gas parcels, accounting for the column density of the absorbers, thermal broadening, and cosmological and Doppler redshifts along the line of sight.  Lastly, it processes the  spectrum to account for instrumental limitations, including Gaussian noise, the spectral resolution and line spread function of the instrument, as well as adding a template quasar spectrum and Milky Way foreground.  The resulting spectrum very closely mimics what is produced by real instruments like the Cosmic Origins Spectrograph (COS) aboard the Hubble Space Telescope (HST).

This was a highly collaborative project,
and accordingly we highlight individual contributions in a footnote at the beginning of each section.

\subsection[Column densities of uniform clouds --- \texttt{sample0}]{Column densities of uniform clouds --- \texttt{sample0}\footnote{
\texttt{sample0} was generated by Cameron Hummels and Zachary Hafen.}}
\label{s: data generation -- sample0}

\begin{figure}
    \centering
    \includegraphics[width=\columnwidth]{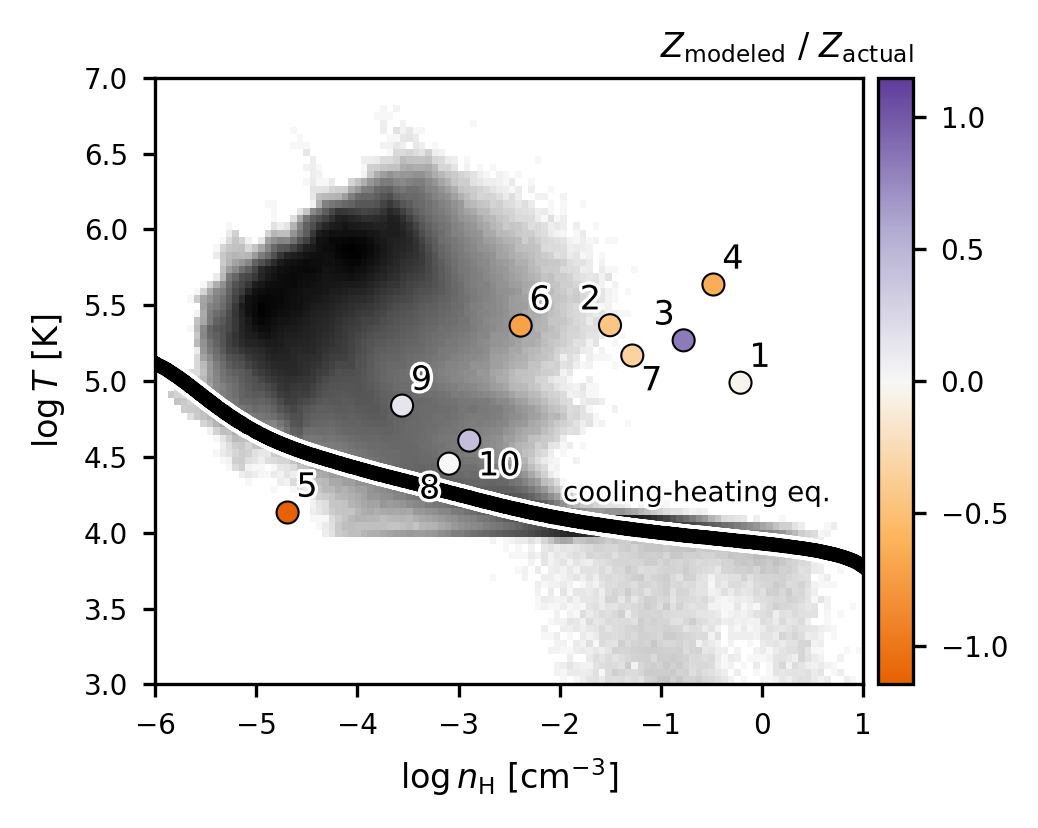}
    \caption{
    Comparison between estimated and actual properties of \texttt{sample0} in the context of the temperature-vs-density distribution of CGM gas in a FIRE-2 cosmological simulation (black background distribution; scales with log of mass and includes satellite ISM).
    The solid black line shows the range of temperatures and densities where gas is in cooling-heating equilibrium via photoheating.
    If cooling-heating equilibrium is assumed during parameter estimation then when the assumption is inaccurate the estimated metallicity tends to be inaccurate.
    However, in the simulation there is a large population of CGM gas that is in approximate cooling-heating equilibrium.
    }
    \label{f: idealized explanation}
\end{figure}
 
The first dataset, \texttt{sample0}, was a highly idealized sample consisting of 10 clouds each with uniform metallicity, density, and temperature.
The data products made available to observers were total  ion column densities for several relevant ions, as opposed to full spectra.
The motivation for \texttt{sample0} was to set a baseline for interpreting the subsequent data samples, absent complications from multiphase structure or interpreting spectra.

The physical properties of the 10 clouds were sampled from $Z=[ 0.01, 1.5 ]~Z_\odot$ (where $Z_\odot = 0.014$ is the total mass fraction in metals; \citealt{asplund2009Chemical}), $n_{\rm H} = [ 10^{-6}, 0.1 ]~{\rm cm^{-3}}$, $T = [ 10^4, 10^6 ]$~K, and $N_{\ion{H}{I}} = [ 10^{15}, 10^{17} ]~{\rm cm^{-2}}$.
The values of density, temperature, metallicity, and \ion{H}{I} column density were chosen to be uncorrelated, which plays a role in the analysis (\S\ref{s: results -- sample0}).
The sampled density, temperature, and metallicity are shown on the left side of Figure~\ref{f: summary}.
We set the redshift for our mock sample to $z=0.25$, typical for many CGM samples observed with the Cosmic Origins Spectrograph (COS, \citealt{green2012COSMIC}), and we included all ions present in the instrument window of COS G130M and G160M at $z=0.25$, including \ion{C}{II}, \ion{C}{III}, \ion{N}{II}, \ion{N}{III}, \ion{N}{V}, \ion{O}{I}, \ion{O}{VI},  \ion{Si}{II}, \ion{Si}{III}, and \ion{Si}{IV}.
For each ion we evaluated the ion densities according to photo-ionization equilibrium, as tabulated by \textsc{trident} for single zone simulations produced with \textsc{cloudy}~\citep{ferland20132013} using the \cite{haardt2012RADIATIVE} UV background.
Note that photo-ionization equilibrium does not assume cooling-heating equilibrium (where gas is in heating-cooling balance, resulting in constant temperature),
an assumption often employed in parameter estimation.
To determine the total amount of absorbing gas given its properties, we calculated the length of the absorber from the \ion{H}{I} column density and its ionization state, $\ell = N_{\ion{H}{}} / n_{\ion{H}{}} = N_{\ion{H}{I}} / n_{\ion{H}{I}}$.
To avoid unphysically large absorbers we discarded absorbers with length $> 1$ Mpc (comparable to a conservative estimate of the diameter of the CGM of a MW-mass galaxy).
Note that the tabulated ion densities used to calculate the ionization state do not assume a size for the cloud, which is acceptable assuming the effects of self-shielding are small.
Using an ionization table that includes self-shielding reveals that self-shielding has a small (typically much less than $\lesssim 0.1$ dex) effect on ion column densities for the clouds considered here.

Prior to making the ion column densities available to the observers, we modified them to reflect uncertainty typical of similar real observations.
The mock errors per cloud per ion, $e_{\log N_{\rm ion}}$, were estimated by sampling a gamma function fit to the distribution of errors estimated as part of the COS-Halos survey~\citep{werk2013COSHALOS}.
Errors a factor of $1.5\times$ beyond the COS-Halos range of errors were re-sampled.
The actual ion column density per cloud per ion made available to observers was then calculated as $\log N_{\rm ion,\,provided} = \log N_{\rm ion} + U(-e_{\log N_{\rm ion}}, e_{\log N_{\rm ion}} )$, where $U$ is a value drawn from a uniform distribution spanning $\pm e_{\log N_{\rm ion}}$.
This is equivalent to assuming that the imaginary observers that observed the synthetic ion column densities calculated a conservative estimate of the error $e_{\log N_{\rm ion}}$, and the real ion column density per cloud per ion lies somewhere within the error bars.

We asked the observers to derive their best estimate for the metallicity, temperature, and density.
Beyond providing the columns and errors on the columns, we informed the observers they may assume each sightline pierced the CGM of a MW-mass galaxy at $z=0.25$, and we suggested using the \cite{haardt2012RADIATIVE} UVB, because of the existence of systematics from the choice of UVB~\citep{Wotta2016, Wotta2019, acharya2021How,Gibson2022}.

\subsection[Spectra of multi-cloud systems --- \texttt{sample1}]{Spectra of multi-cloud systems --- \texttt{sample1}\footnote{
\texttt{sample1} was generated by Nastasha Wijers, Zachary Hafen, and Cameron Hummels.
}}
\label{s: data generation -- sample1}

\begin{figure*}
    \centering
    \includegraphics[width=\textwidth]{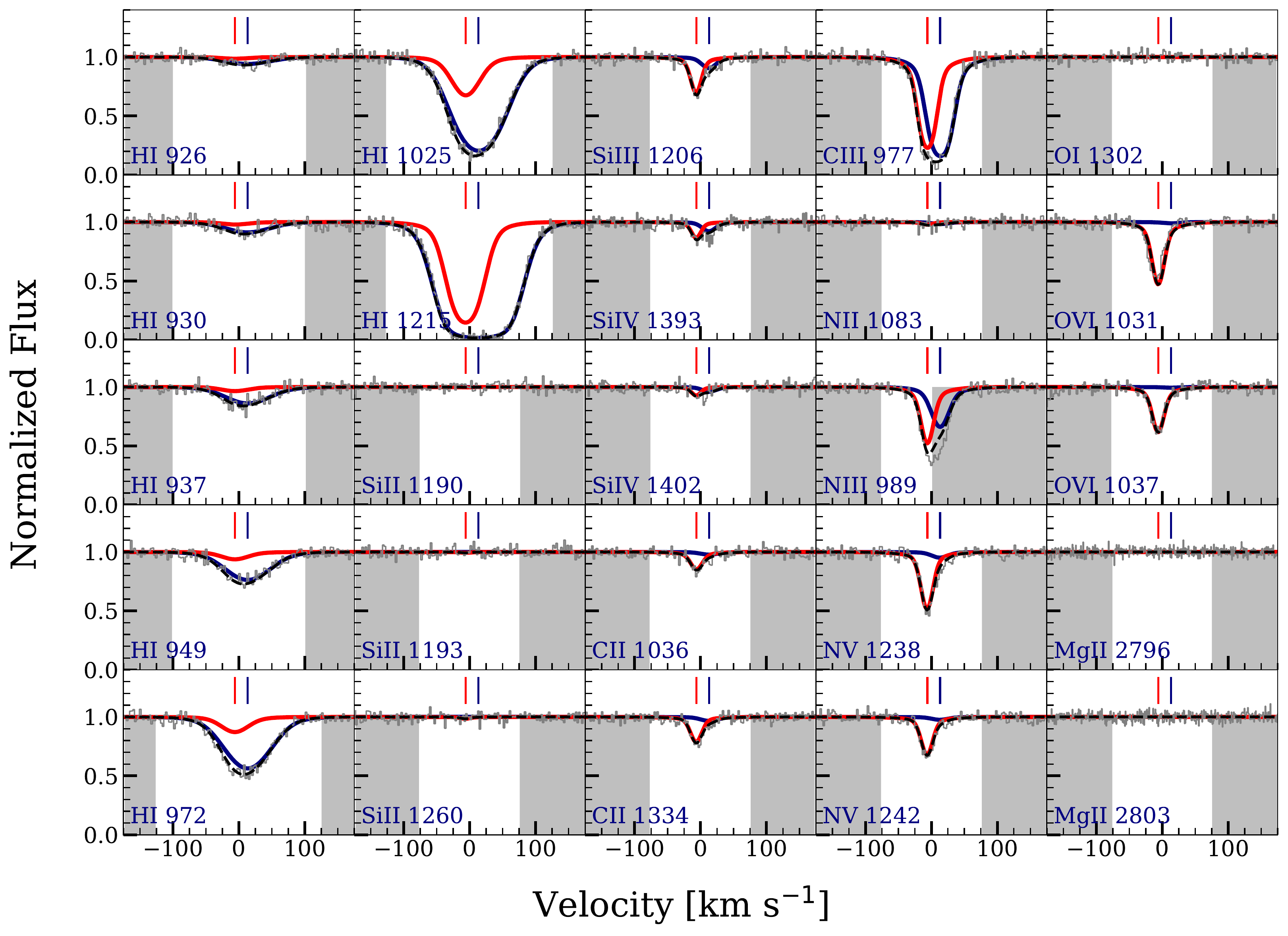}
    \caption{
    The source spectrum (grey steps) and the parameter-estimation best-fit spectrum (dashed black) for sightline 076 in \texttt{sample1}.
    The data are best fit by a two-cloud model, and the Voigt profiles of the individual components are shown in dark blue and red. 
    The dark blue curve traces the phase that produces the bulk of the \ion{H}{I} absorption, and the red curve traces the phase that produces \ion{O}{VI} absorption.
    The spectrum wavelength was converted to line-of-sight velocity in this image,
    and the velocity of each component is indicated by a vertical dash.
    The grey bands show the portion of the spectrum with no influence on parameter estimation.
    }
    \label{f: sample1 spectrum}
\end{figure*}

\begin{figure}
    \centering
    \includegraphics[width=\columnwidth]{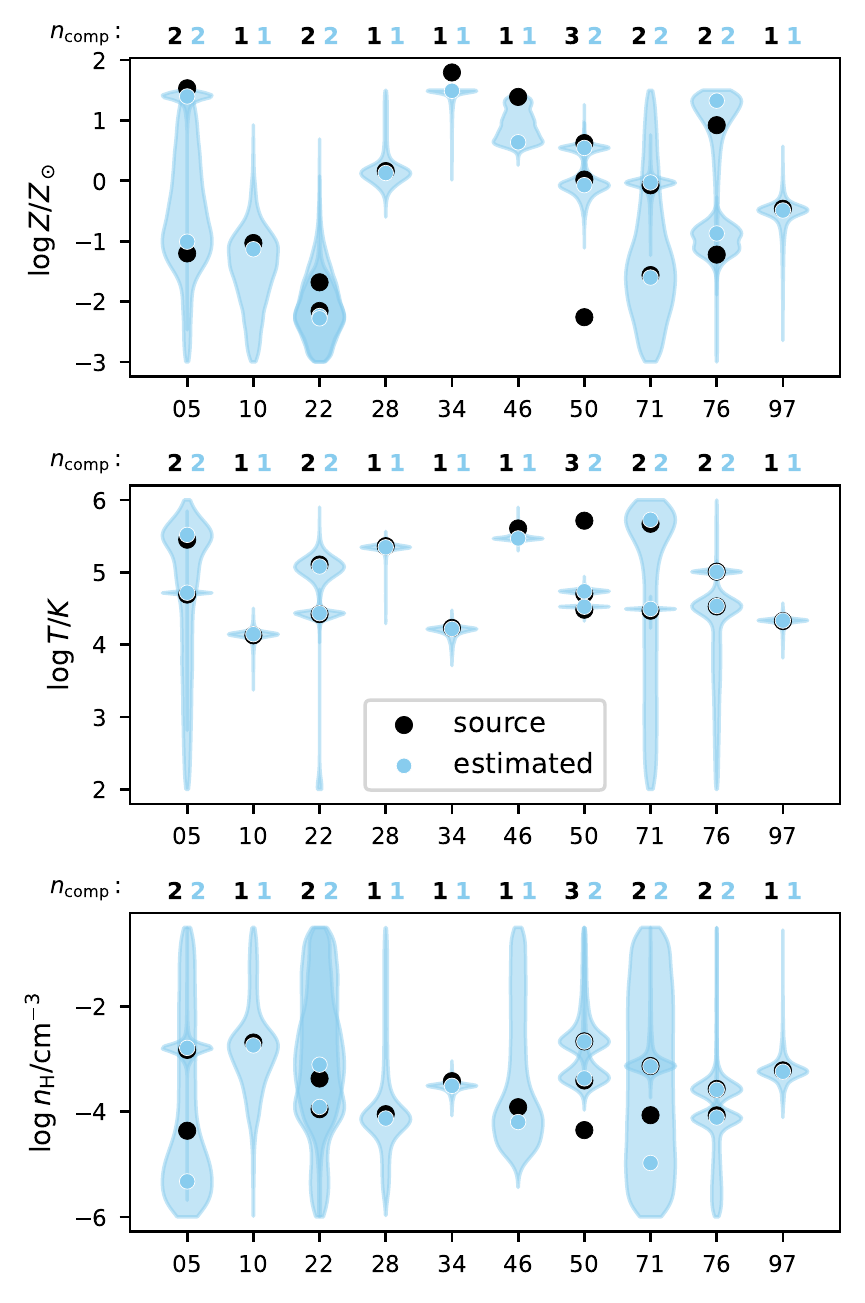}
    \caption{
    Synthetic data properties for multiple uniform clouds (\texttt{sample1}, black) compared to the distribution of expected values (the posteriors) from parameter estimation (blue).
    The wider the ``violin'' plots, the larger the value of the posterior at that value.
    As an example, for sightline 05 the parameter estimation suggests two components with metallicities \metallicity $\sim [ 1, -2]$ and \metallicity $\sim [ 0.5, 1.5 ]$.
    The numbers above each sightline indicate the actual number of source clouds compared to the estimated number of clouds.
    The maximum likelihood estimates (the peak values of the posteriors) typically align with the source parameters.
    }
    \label{f: sample1 violin}
\end{figure}

This sample was designed to test how well multiple clouds could be modeled.
From \texttt{sample0} to \texttt{sample1} we transitioned from providing only total ion column densities to providing full spectra generated with \textsc{trident} (shown in Figure~\ref{f: sample1 spectrum}).
We also moved from one discrete, uniform cloud per sightline to up to three discrete, uniform clouds per sightline.
We used KDE resampling with \textsc{kalepy}~\citep{kelley2021Kalepy} to draw the properties of the clouds from the distribution of properties typical for the CGM in the \textsc{EAGLE} cosmological simulations.
The sampled metallicity, density, and temperature are shown in Figure~\ref{f: sample1 violin}.

EAGLE \citep[`Evolution and Assembly of GaLaxies and their Environments';][]{schaye2015EAGLE,Crain2015,McAlpine2016} is a cosmological, hydrodynamical simulation.
It uses the Gadget3 ``Tree-PM'' scheme for gravity \citep[][]{springel2005Cosmological} and the Anarchy implementation of smooth particle hydrodynamics (SPH) for fluid calculations (\citeauthor{schaye2015EAGLE} \citeyear{schaye2015EAGLE}, appendix~A; \citeauthor{Schaller2015} \citeyear{Schaller2015}).
The Ref-L100N1504 simulation, used here, was run in a $(100\,\mathrm{cMpc})^3$ periodic volume, with a mass resolution of $9.70 \times 10^6 \,\mathrm{M}_{\odot}$ for dark matter, an initial SPH particle mass of $1.81 \times 10^6 \,\mathrm{M}_{\odot}$, and a Plummer-equivalent gravitational softening length of $0.70$~pkpc (at low redshift).
In addition to gravity and hydrodynamics, the EAGLE simulations include radiative cooling and heating \citep{wiersma2009Effect}, star formation \citep{Schaye2004, Schaye2008}, stellar feedback \citep{DallaVecchia2012}, and metal enrichment by stars \citep{wiersma2009Chemical}, as well as black hole seeding, growth \citep{Rosas-Guevara2015}, and AGN feedback \citep{Booth2009}.
For more details, see the EAGLE core papers~\citep{schaye2015EAGLE,Crain2015,McAlpine2016}.

To generate the distribution from which to sample the cloud properties we selected 200 random galaxies with halo masses $M_{200c} = 10^{12} - 10^{12.5} M_\odot$ from the EAGLE Ref-L100N1504 volume at $z=0.27$, where $M_{200c}$ is the halo mass such that the average enclosed density is 200$\times$ the critical density of the universe at that redshift.
In the gas around these galaxies, the \ion{H}{I} fraction was calculated using fitting functions that approximate the effects of self-shielding~\citep{rahmati2013Impact} assuming the UVB from \citet{haardt2001Modelling}.
We extracted 200 sub-volumes around the 200 galaxies, and in each sub-volume we performed an \ion{H}{I}-mass-weighted projection of all gas with impact parameter $< R_{200c}$ and LOS distance $< 2 R_{200c}$ relative to the galaxy. The projection was done along the z-axis of the simulation, which is random with respect to the galaxy orientation.
For each halo, we obtained  `maps' of \ion{H}{I} column density and \ion{H}{I}-weighted temperature, density, and metallicity, with a pixel area of (3.125 comoving kpc)$^2$.
This corresponds to $\sim 10,000$ sightlines for a MW-mass galaxy.
We then created a histogram of these sightline properties (simply weighted by sightline count) to try to capture a physically reasonable joint distribution of temperature, density, metallicity, and \ion{H}{I} column density in UV absorbers.
Note that EAGLE does not include molecular cooling, so temperatures below $10^4$ K are rare, and the temperature of dense gas ($\mathrm{n}_{\mathrm{H}} > 10^{-1} \, \mathrm{cm}^{-3}$) is set by an equation of state and is therefore not physically realistic \citep{schaye2015EAGLE}.
Therefore, when calculating the hydrogen neutral fraction we imposed a temperature of $10^4$~K on the gas on this equation of state~\citep[following e.g.,][]{Rahmati2016}.

We generated 100 sightlines of one to three clouds selected from the above distribution. We set $v_{\rm LOS}$ for each cloud to be within $\pm150$ {\kms} of $z=0.25$, and randomly sampled temperature, density, metallicity, and $N_{\ion{H}{I}}$ from the cosmological distribution.
To focus on interpreting sightlines with multiple cool clouds we only allowed a single cloud per sightline to have $T>10^5$ K.
We determined the ion densities and column densities as described for \texttt{sample0}, using an ionization table and constraining total column via $N_{\ion{H}{I}}$.
We used \textsc{trident} to produce spectra for both COS G130M and COS G160M, and we also provided spectra for the \ion{Mg}{II} doublet at $\lambda = 2796, 2803$~{\AA} to mimic complementary observations from the ground with a resolving power of $\approx$ 40,000.
The $\alpha$ element abundances relative to other elements were set to solar, e.g. \ion{C}{}/$\alpha$ is solar.
We applied the COS line spread function to each spectrum and included Gaussian noise with an signal-to-noise ratio of 30.
Data with this low noise are unusual (as discussed in \S\ref{s: discussion -- cloud structure -- recovery}), so this sample provides a baseline for performance in a best-case scenario.
The wavelength resolution for each spectrum was set to $0.01$ \AA.
While we generated 100 sightlines, it is challenging to model that many.
Accordingly observers sampled 10 random sightlines to focus on, excluding any damped Ly $\alpha$ absorbers to mitigate the effect of self-shielding.
The only additional information provided to observers was that they could assume the absorbers were found in a galaxy-targeted survey with the host galaxies located at $z=0.25$.

\subsection[Spectra of high-resolution turbulent mixing --- \texttt{sample2}]{Spectra of high-resolution turbulent mixing --- \texttt{sample2}\footnote{
\texttt{sample2} was generated by Nir Mandelker, Zachary Hafen, and Cameron Hummels.}}
\label{s: data generation -- sample2}

\begin{figure*}
    \centering
    \includegraphics[width=0.49\textwidth]{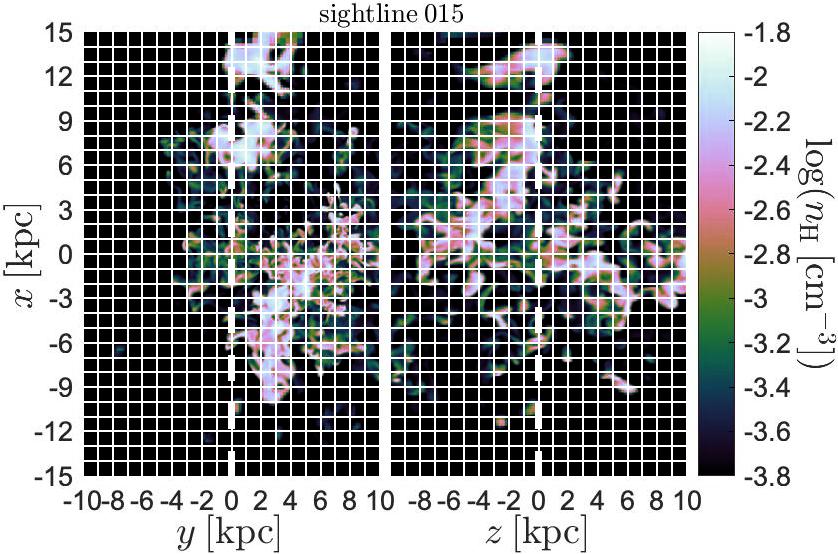}
    \includegraphics[width=0.49\textwidth]{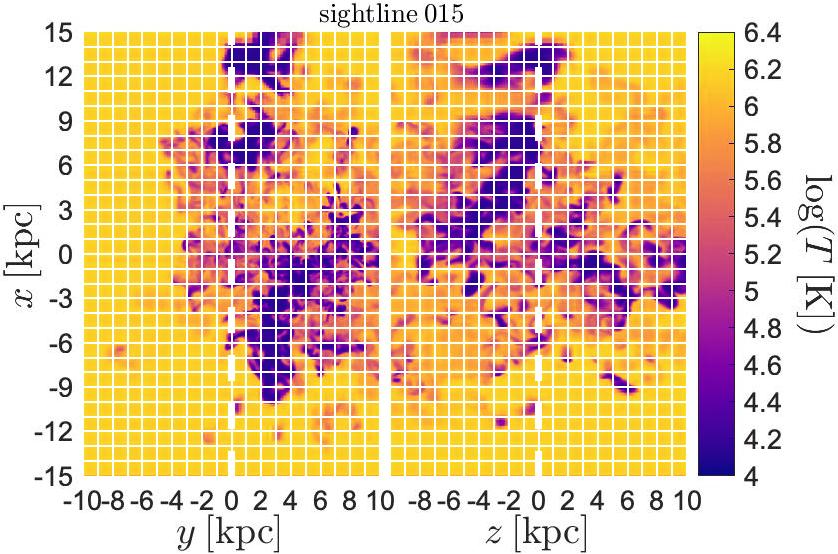} \\
    \includegraphics[width=0.49\textwidth]{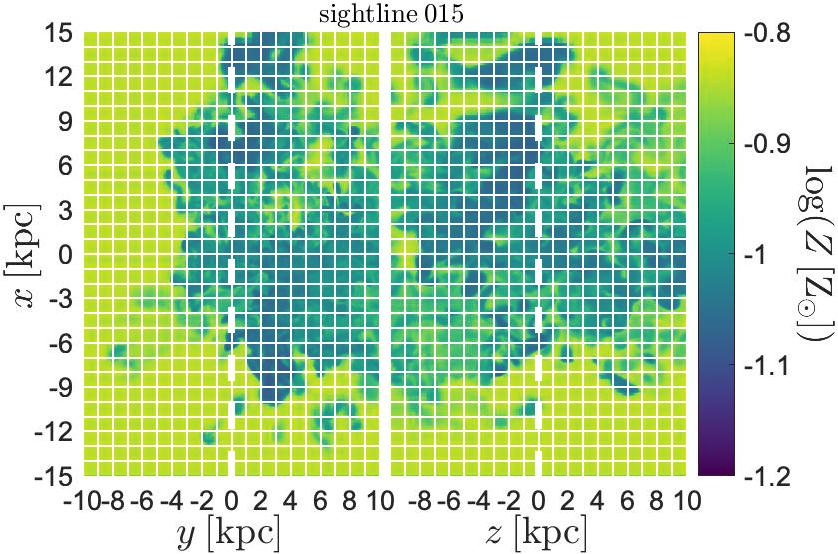}
    \includegraphics[width=0.49\textwidth]{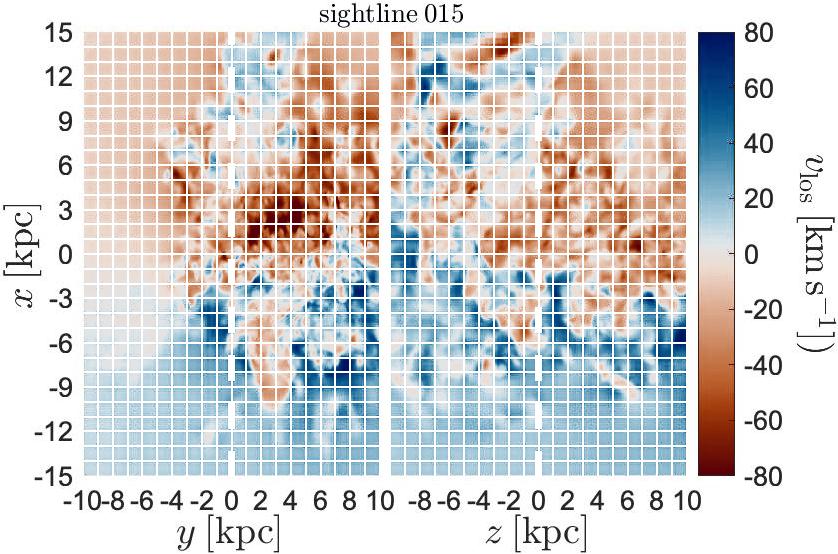}
    \caption{
    Density, temperature, metallicity, and line-of-sight velocity in a slice of the simulation used to generate \texttt{sample2}~\citep{mandelker2020Instability}.
    The dashed white line shows the location of one of the sightlines (sightline 15) forward-modeled to produce mock spectra.
    The simulation includes a turbulent mixing zone, and shows highly complex cloud structure.
    A grid with 1 kpc spacing is overlaid in white to help the viewer gauge cloud size.
    }
    \label{f: sample2 ray 15}
\end{figure*}

\begin{figure*}
    \centering
    \includegraphics[width=\textwidth]{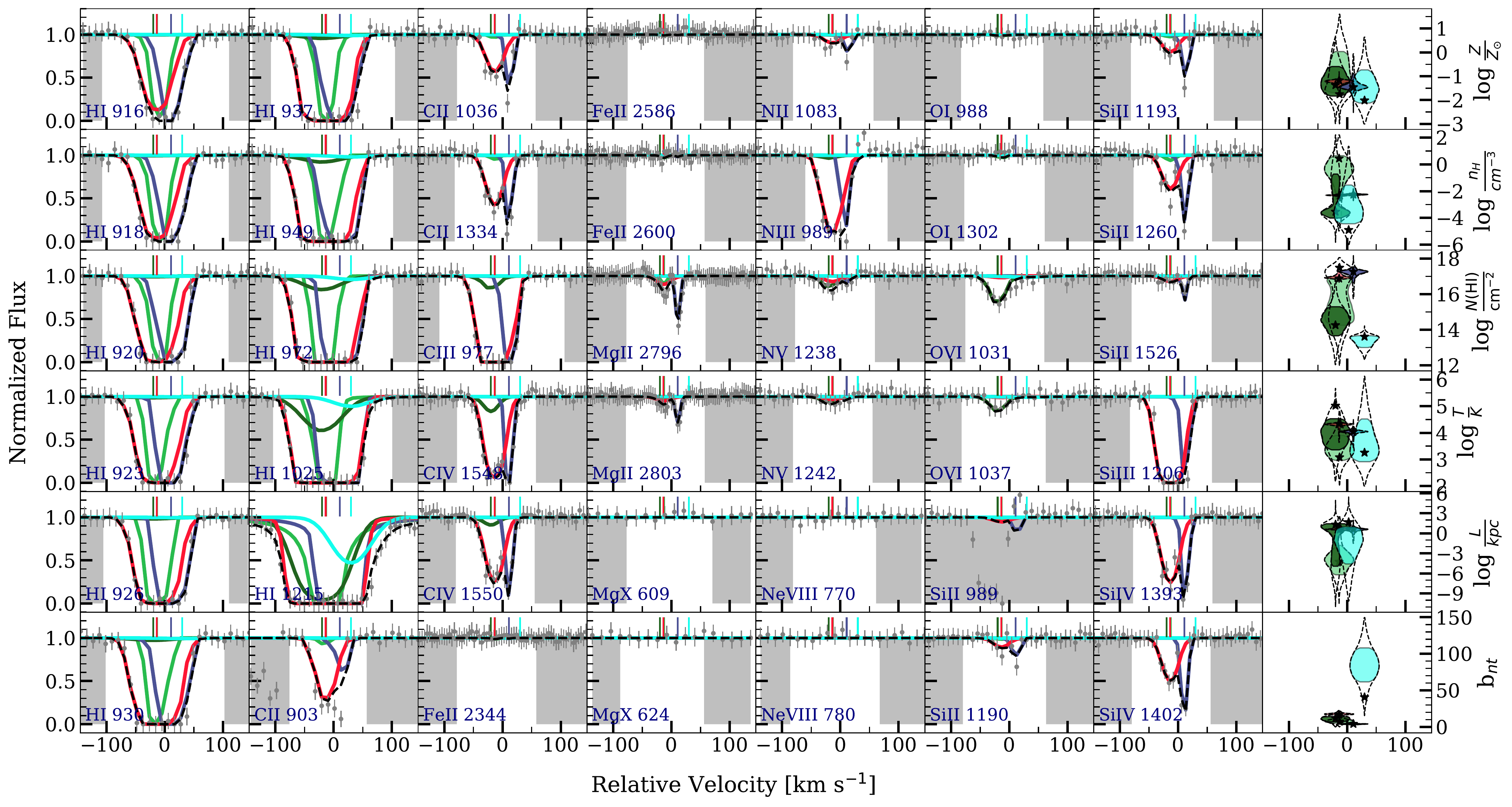}
    \caption{
    Mock spectra for the sightline shown in Figure~\ref{f: sample2 ray 15},
    best fit absorption profiles found by observers,
    and the parameters of the best fits.
    The absorption along this sightline is best fit by five components.
    The stars in the right panel indicate the maximum likelihood estimate (MLE) according to best fit across all iterations tested during parameter estimation,
    which is subject to noise.
    The MLE displayed elsewhere in the analysis traces the peak of the posterior distribution.
    }
    \label{f: sample2 spectrum 15}
    \end{figure*}

\begin{figure*}
    \centering
    \includegraphics[width=\textwidth]{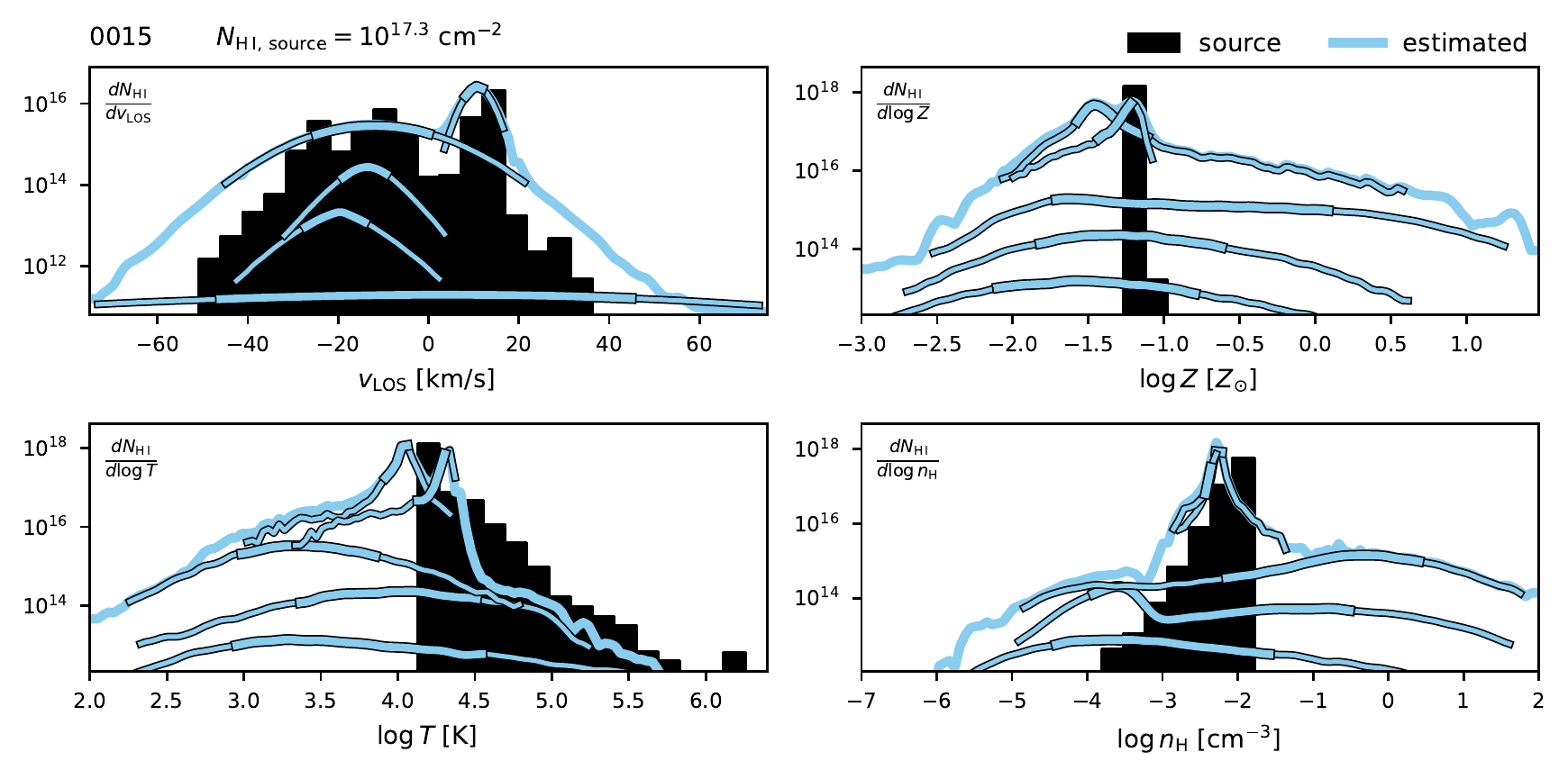}
    \caption{
    Estimated and source properties of gas along the sightline shown in Figures~\ref{f: sample2 ray 15} and \ref{f: sample2 spectrum 15}.
    Black shows the source properties used to produce the mock spectra.
    Blue shows the posteriors from the parameter estimation. 
    The combined posteriors are indicated by thick lines.
    We also show the posteriors for individual clouds as thick (thin) blue lines outlined in black, for the regions enclosing 68\% (99.7\%) of the posterior for a given cloud.
    The clouds that dominate \ion{H}{I} absorption agrees with the peaks of the source data, while additional clouds have poorly constrained parameters.
    }
    \label{f: sample2 15}
\end{figure*}

\begin{figure}
    \centering
    \includegraphics[width=0.95\columnwidth]{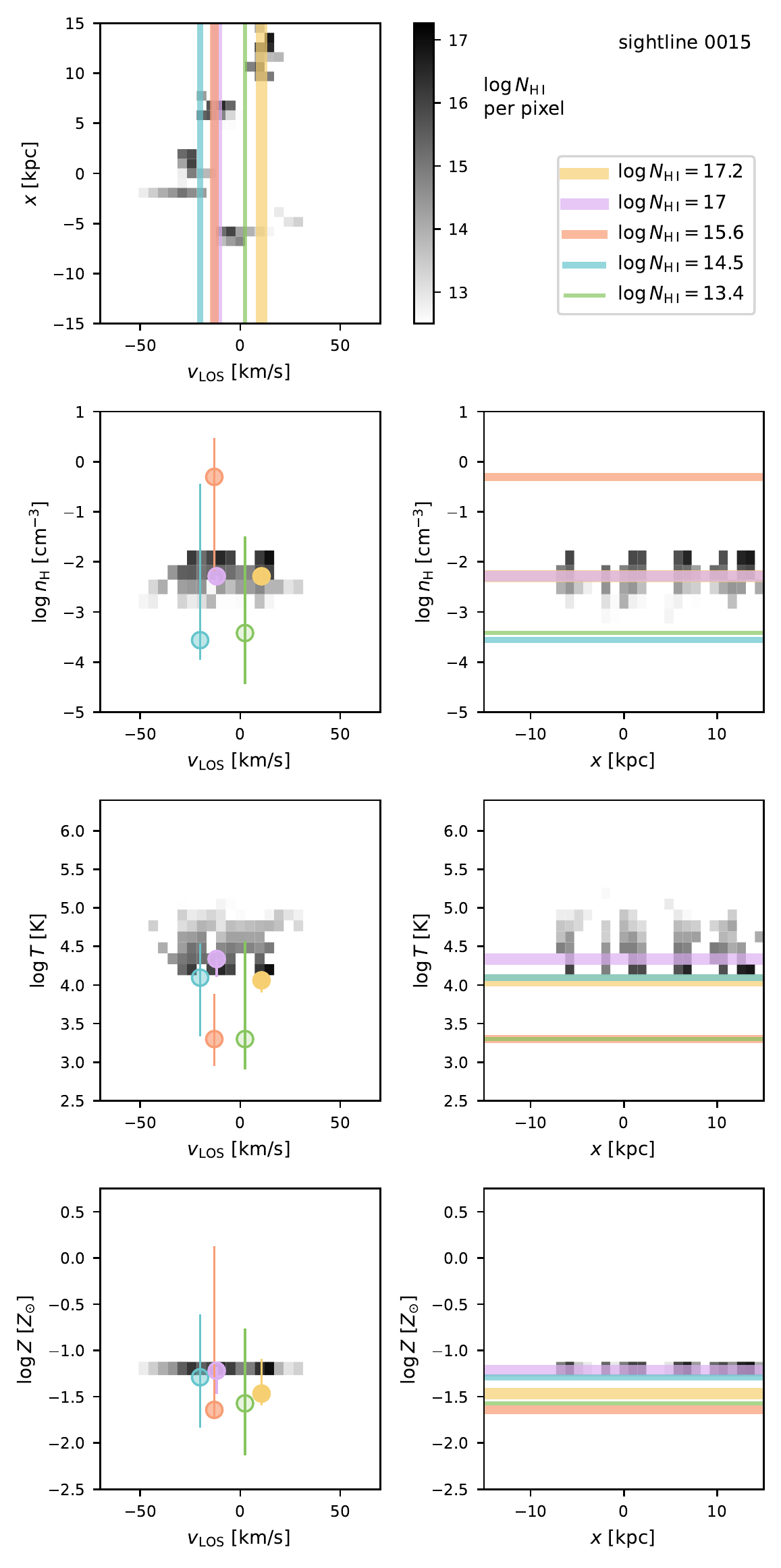}
    \caption{
    Physical- and velocity-space cloud structure for the sightline shown in Figures~\ref{f: sample2 ray 15}, \ref{f: sample2 spectrum 15}, and \ref{f: sample2 15}.
    The top left panel shows the relationship between position along the line of sight ($x$) and the line-of-sight velocity of the gas.
    Source properties are plotted with black 2D histograms,
    while the colors correspond to estimated components.
    The estimated components accurately describe the gas responsible for the vast majority of the absorption.
    However, gas aligned in velocity but physically separated can introduce confusion to the interpretation,
    especially when structure beyond the dominant absorption is loosely constrained.
    }
    \label{f: sample2 structure 15}
\end{figure}

\begin{figure*}
    \centering
    \includegraphics[width=\textwidth]{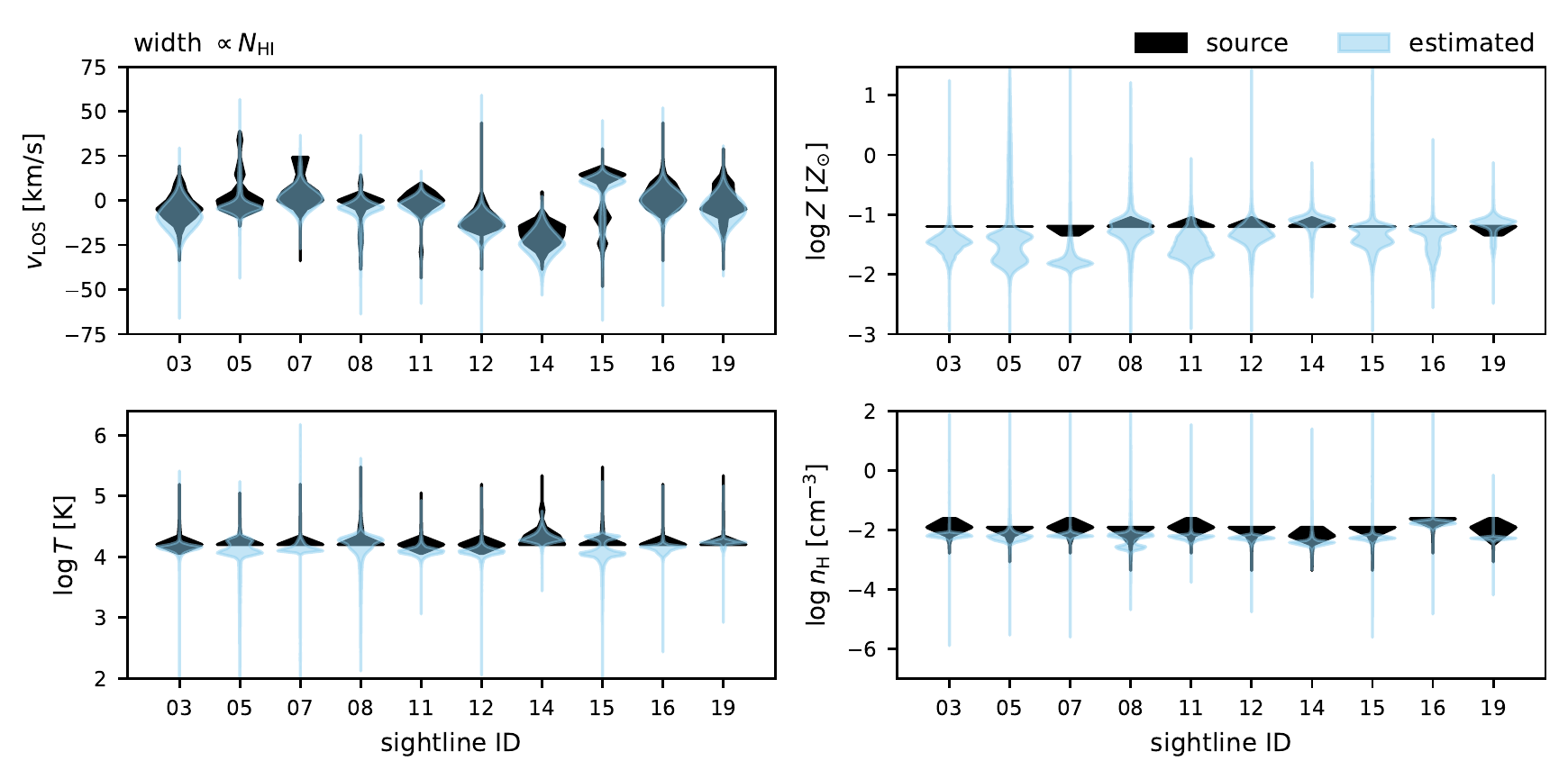}
    \caption{
    The distributions of $v_{\rm LOS}$, $Z$, $T$, and $n_{\rm H}$ for the synthetic source data (black) compared to the full posteriors from parameter estimation (blue).
    The width of the violin plots scales with $N_{\ion{H}{I}}$.
    The parameter estimation accurately describes the properties of the gas responsible for the majority of the absorption.
    }
    \label{f: sample2 violin}
\end{figure*}

\begin{figure*}
    \centering
    \includegraphics[width=\textwidth]{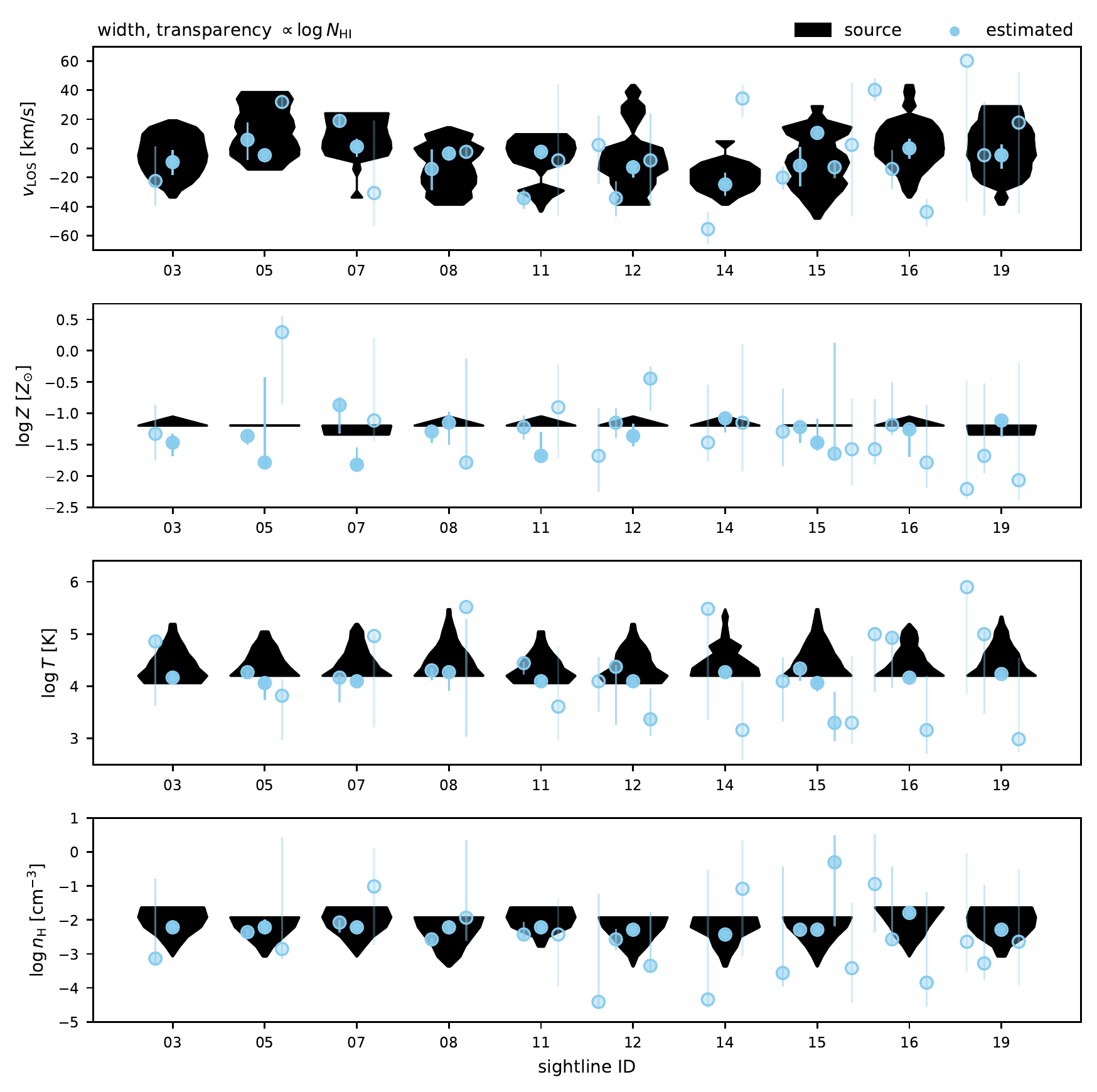}
    \caption{
    The distributions of $v_{\rm LOS}$, $Z$, $T$, and $n_{\rm H}$ for the synthetic source data (black) compared to the best estimates for individual components comprising the parameter estimation.
    Each point is the maximum-likelihood estimate (MLE) of an individual component,
    and the associated bars enclose the 16th to 84th percentiles of the posterior.
    The width of the violin plots and the opacity of the points are scaled \textit{logarithmically} with $N_{\ion{H}{I}}$,
    such that if the parameter estimation perfectly agreed with the data all visible MLE points would lie within the source data violins.
    As shown above, the strongest component(s) traces the average absorbing gas properties.
    The lower-$N_{\ion{H}{I}}$ components trace some features of the underlying distribution,
    but arguably the lower-$N_{\ion{H}{I}}$ structure present in the source data is not retrieved.
    }
    \label{f: sample2 violin vs components}
\end{figure*}

We designed the final sample to test modeling absorption systems with a realistic multiphase \textit{distribution} of properties, as opposed to being described by a few discrete clouds.
This choice reflects growing knowledge that instabilities may drive clouds to fragment and mix with surrounding gas (see \citealt{faucher-giguere2023Key} for a review).
For \texttt{sample2} we used data from a high-resolution simulation of a cool filament embedded in a hot halo~\citep{mandelker2020Instability}.
A turbulent mixing layer forms at the boundary of the filament and the hot gas, which produces gas with a complex continuous distribution of temperature, density, and metallicity.
Figure~\ref{f: sample2 ray 15} shows an example sightline.

The simulation used here is the M1.0\_d100 simulation from Table 1 of \cite{mandelker2020Instability}.
This simulation uses the AMR code \texttt{RAMSES} \citep{teyssier2002Cosmological} to simulate a cold, dense cylindrical stream flowing supersonically through a hot, diffuse background, initially in pressure equilibrium.
This represents a cold stream tracing a cosmic-web filament as it flows towards the central galaxy through its hot CGM, predicted to be the main mode of gas accretion in massive high-$z$ galaxies~(\citealt{keres2009Galaxies, dekel2009Cold}, though see~\citealt{nelson2013Moving}).
The initial density in the stream is $n_{\rm H,s}=10^{-2}$ cm$^{-3}$ and the density in the background is smaller by a factor of $\delta=100$, $n_{\rm H,b}=10^{-4}$ cm$^{-3}$.
The temperature in the cold stream is set by cooling-heating equilibrium with a $z=2$ \cite{haardt1996Radiative} UV background with no self-shielding for dense gas, resulting in $T_{\rm s}\sim 1.5 \times 10^4$ K.
The hot and cold gas are initially in pressure equilibrium, which sets the temperature in the hot phase, $T_{\rm b}\sim 1.5 \times 10^6$ K.
Note that at this temperature there is insufficient rest-frame UV absorption to accurately model the hot phase, which affected the parameter estimation.
The stream is initialized with a metallicity of $0.03 Z_\odot$ (with $Z_\odot = 0.02$), while the background is initialized with a metallicity of $0.1 Z_\odot$.
The individual abundances are set to solar composition, a fiducial choice for many simulations.
The stream initially flows parallel to its axis with a velocity equal to the sound speed in the hot background.

The stream radius is set to $R_{\rm s}=3$ kpc, and the simulation domain is a cube with sides $L=32R_{\rm s}$, extending from $-16R_{\rm s}$ to $16R_{\rm s}$ in all directions, with the stream axis corresponding to the $z$-axis.
We use periodic boundary conditions at $z=\pm 16R_{\rm s}$ and outflow boundary conditions at $x=\pm 16R_{\rm s}$ and $y=\pm 16R_{\rm s}$, such that gas crossing the boundary is lost from the simulation domain.
At these boundaries, the gradients of density, pressure, and velocity are set to 0.
The stream fluid initially occupies the region $r<R_{\rm s}$ while the background fluid occupies the rest of the domain.
We use a statically refined grid with resolution decreasing away from the stream axis.
The resolution is highest in the region ${\rm max}(|x|,|y|)<1.5R_{\rm s}$.
Cells that lie within this region have a cell size $\Delta=R_{\rm s}/64\approx 47$ pc.
The cell size increases by a factor of 2 every $1.5R_{\rm s}$ in the $x$ and $y$-directions, up to a maximal cell size of $\Delta=R_{\rm s}/4=0.75$ kpc.
The resolution is uniform along the $z$ direction, parallel to the stream axis.
As shown in \cite{mandelker2020Instability}, this resolution is sufficient to reach convergence in the evolution of the stream-background interaction.

As the simulation progresses, Kelvin-Helmholtz Instability (KHI) develops at the interface between the stream and the background, forming a radiative turbulent mixing layer around the stream where the two phases mix.
For the chosen parameters, the cooling time in the mixing layer is shorter than the Kelvin-Helmholtz mixing timescale \citep{mandelker2020Instability}.
As a result, initially hot CGM gas cools and condenses onto the stream through the mixing layer, and is entrained in the flow, causing the cold gas mass to increase with time.
The snapshot used for spectra in this work is 10 stream sound crossing times into the evolution of the stream.
At this time the stream is largely fragmented into small clouds, with significant mixing between phases.

To generate spectra we first used cloud-in-cell interpolation to deposit onto 20 sightlines the volume-weighted density and mass-weighted temperature, metallicity, and LOS velocity.
The binning along each sightline was set to twice the smallest cell in the simulation, $2\Delta \approx 100{\rm pc}$.
We selected the sightlines to maximize the variation in density, temperature, and LOS velocity along each sightline.
All sightlines intersected the turbulent mixing zone at the boundary of the filament and the hot CGM, with five of them intersecting both the turbulent mixing zone and the filament core.
All 20 sightlines extend half the box length and are parallel to the $x$ direction.
These are divided into four groups of five sightlines each, which intersect the $x=0$ plane at four different $z$ coordinates along the filament length.
The five sightlines in each such group intersect the $x=0$ plane at five different $y$ values, extending from $\pm 3.2R_{\rm s}$, thus probing the turbulent mixing zone at different distances from the filament axis, with one sightline in each group passing through the filament core at $y=0$.
We generated spectra for all twenty sightlines, all of which show clear multiphase structure in the spectra for multiple ions.
Of the twenty sightlines we provided ten to observers, chosen to span the breadth of output spectra.
An example sightline is shown in Figure~\ref{f: sample2 ray 15} and a summary of $v_{\rm LOS}$, $Z$, $T$ and $n_{\rm H}$ for the sightlines is shown in Figure~\ref{f: sample2 violin vs components}.

For the blinded data sample we changed the redshift to $z=0.13$ to bring \ion{C}{IV} into the observable range for the COS gratings.
The simulations were run including a $z=2$ UVB.
However, during the initial analysis the ionization fractions were calculated in post-processing via Trident using a $z=0.13$ \cite{haardt2012RADIATIVE} UVB, and the observers were told to use the same UVB.
The motivation for producing $z=0.13$ spectra from a simulation run at $z=2$ was two-fold:
to avoid providing any clues regarding the origin of the blinded data,
and to test the extent to which parameter estimation is successful at extracting gas properties independent of the mechanisms responsible for producing gas with a given $Z$, $T$, and $n_{\rm H}$.
For this sample, following the initial analysis we provided the observers with revised synthetic data.
The revised data used a $z=2$ UVB for calculating the ionization fractions, and was ``observed'' at $z=2$.
The revised spectra had a wavelength resolution of 0.05 \AA, a line spread function with a FWHM of 7 {\kms}, and included the ions \ion{H}{I}, \ion{Si}{II}, \ion{Si}{III}, \ion{Si}{IV}, \ion{C}{II}, \ion{C}{III}, \ion{C}{IV}, \ion{O}{I}, \ion{O}{VI}, \ion{N}{II}, \ion{N}{V}, \ion{Fe}{II}, \ion{Mg}{X}, and \ion{Ne}{VIII}.

\textsc{trident} uses absorption-line properties archived by the National Institute for Standards and Technology as a default, and we kept this default for the first two samples.
For the third sample we aimed for greater consistency with the parameter estimation methodology by retrieving absorption line information from the commonly-used spectra analysis package \textsc{linetools}~\citep{prochaska2016Linetools}.
Between these two datasets the oscillator strengths could disagree by up to $\sim 2$ dex for the lines \ion{O}{I} 922, 925, 930, and 989, \ion{N}{I} 1200, \ion{Si}{II} 1304, and \ion{S}{IV} 1073.
In the future the default behavior for \textsc{trident} will be to use the \textsc{linetools} dataset.

\subsection[Parameter estimation]{Parameter estimation\footnote{
 Parameter estimation was performed by Sameer and Jane Charlton.}}
\label{s:  parameter estimation}

 Parameter estimation employed the cloud-by-cloud, multiphase, Bayesian ionization modeling (CMBM) methods introduced in \cite{sameer2021Cloudbycloud}, and further refined in \cite{sameer2022Probing}.
Previous work demonstrated that the metallicity can vary substantially in velocity space across an observed absorption line system~\citep{Prochter2010,lehner2019COS,Wotta2019,Zahedy2021,lehner2022Intermediate}, and thus cloud-by-cloud modeling is important.
As with the synthetic data generation, the parameter estimation used \textsc{cloudy}~\citep{ferland20132013} with the \cite{haardt2012RADIATIVE} UVB.

For \texttt{sample0}, the initial modeling procedure involved adjustment of the three free parameters:
\ion{H}{I} column density ($N_{\ion{H}{I}}$), metallicity ($Z$), and hydrogen number density ($n_{\rm H}$).
For each point in parameter space we used \textsc{cloudy} to calculate the temperature based on a balance of heating and cooling,
and determined the column densities of all ions provided by the simulators, assuming photoionization equilibrium.
A log-likelihood function was computed by summing the chi-squared between the provided and the model column densities for all ions, accounting for the provided errors for each column density.
This function was minimized to determine the distribution of the three parameters.
The parameter estimation was performed using PyMultiNest~\citep{buchner2016Statistical}. 

For the revised parameter estimation of \texttt{sample0} we also added in the temperature, $T$, as a free parameter, no longer assuming a balance of heating and cooling.
At temperatures significantly higher than $T_{eq}$ (see thick curve in Figure~\ref{f: idealized explanation}), the gas is predominantly collisionally ionized.
The distribution of the four parameters, $N_{\ion{H}{I}}$, $Z$, $n_{\rm H}$, and $T$, was then evaluated as constrained by the model column densities.
\S\ref{s: discussion -- modeling choices} contains further discussion regarding equilibrium assumptions.

For the other two samples, \texttt{sample1} and \texttt{sample2}, model spectra were provided.
The shapes of the model line profiles provide an important constraint using the methods of \cite{sameer2021Cloudbycloud} and \cite{sameer2022Probing}.
Though the  parameter estimation methods employed by \cite{zahedy2019Probing} and \cite{haislmaier2021COS} use a cloud-by-cloud approach, they do not use profile shapes, but instead use measured column densities.
The shapes of the profiles, as employed here, give more leverage to separate phases that are coincident in velocity.
The important transitions that were covered by the provided  data included \ion{Mg}{II}, \ion{C}{II}, \ion{C}{III}, \ion{Si}{II}, \ion{Si}{III}, \ion{Si}{IV}, \ion{N}{V}, and \ion{O}{VI} for \texttt{sample1}, and additionally \ion{C}{IV}, \ion{Ne}{VIII}, and \ion{Mg}{X} for \texttt{sample2}.
The given spectra are inspected for absorption in various ionization transitions, and an initial VP fit was obtained for representative unsaturated transitions for a range of ionization states.
The VP fit to the low ionization transitions is used to set a starting number of component locations for the modeling, and their initial positions in velocity space.

CLOUDY models are generated over the four-parameter space defined by $N_{\ion{H}{I}}$ = [10$^{10}$, 10$^{21}$] cm$^{-2}$, $Z$ = [10$^{-3}$, 10$^{1.5}$] $Z_\odot$, $T$ = [10$^{2}$, 10$^{6.5}$] K, and $n_{\rm H}$ = [10$^{-6}$, 10$^{2}$] cm$^{-3}$.
For the set of initial model clouds, a spectrum is synthesized using the superposition of all clouds.
The VP for each cloud comes from the CLOUDY output at a given grid point, a Doppler parameter which is the sum of contributions of the thermal component, and a non-thermal Doppler parameter which is a fifth free parameter of the model.
The velocities of each of the clouds are also allowed to vary, making the redshift a sixth parameter for each cloud.
The grid of CLOUDY models is then explored for the combination of the multiple clouds to determine the parameters for which the log-likelihood function is minimized.
In this case, the log-likelihood function is computed by comparing the model spectra to the provided spectra, and summing the differences pixel by pixel.
If the starting number of component clouds from the low ionization VP fits do not completely account for the absorption in the higher ionization transitions, then additional components are included and the procedure is repeated.
It is found, for example, that \ion{O}{VI} seldom arises in the same phase as \ion{Mg}{II}.
The posterior distribution of parameters for each of the clouds is then determined, again using PyMultiNest.

\cite{liang2018Observing} used simulated spectra to show that Bayesian ionization modeling techniques could estimate properties in close agreement with the \ion{H}{I} mass-weighted averages along the LOS.
This method follows earlier work~\citep{crighton2015Metalenriched, fumagalli2016Physical}, and additionally performs a careful treatment of non-detections.
However, it assumes that the column density constraints for different ions are independent of each other.
This presumption compels each component's observed column density to originate from a single gas cloud parcel.

However, multiphase gas may have contributions from one or more gas clouds in a single component.
The CMBM approach enables more-precise constraints by using the shapes of the absorption profiles rather than only the measured component column densities.
By accounting for their position in velocity relative to the Lyman series absorption, metallicities of different clouds are constrained.
The constraint comes from the closest edge of the \ion{H}{I} profile if a metal-line component is not centered on the \ion{H}{I} profile.
Given that the $b$ parameters of the metal lines and the \ion{H}{I} must be self-consistent, taking into account the derived non-thermal $b$ from the optimization, the CMBM method is able to extract detailed information from the absorption profiles.

\section{Results}
\label{s: results}

We discuss the per-sample results in the following subsections, while Figure~\ref{f: summary} summarizes some of the key results.
The overlap of points in Figure~\ref{f: summary} reflects the agreement between the parameter-estimation best estimates and the source properties.
For \texttt{sample0} and \texttt{sample1} we show one value per cloud for both the source and parameter estimations.
For \texttt{sample2} the source data is a distribution of clouds, and therefore we show the 16th to 84th percentiles of the \ion{H}{I}-weighted distributions compared to the \ion{H}{I}-weighted maximum likelihood estimates from parameter estimation.
The parameter estimates and the source properties agree well across all three samples,
and when there are disagreements, the estimated uncertainty is typically sufficiently conservative.

\subsection{Column densities of uniform clouds --- \texttt{sample0}}
\label{s: results -- sample0}

The left side of Figure~\ref{f: summary} shows the results of parameter estimation for \texttt{sample0}.
The blinded estimates (red)  are the observers' first estimates given the data and no additional information.
After the observers compared the blinded estimates to the revealed source properties they made adjustments to their methodology and performed parameter estimation again.
The results are the revised estimates (blue).
For \texttt{sample0} the primary revision to methodology was removing the assumption of cooling-heating equilibrium.
In addition, in the revision the same procedure was applied to all sightlines, instead of adapting the methodology for each sightline, as was done for the blind estimates.

The discrepancies between the blinded estimated properties and the source properties are driven by the assumptions made by both observers and theorists.
Theorists created \texttt{sample0} by randomly sampling properties from a uniform distribution for each property, assuming for simplicity the properties were independent.
Subsequently, in isolation each property was characteristic of CGM gas, but in combination the properties described gas atypical for the CGM.
This is seen in Figure~\ref{f: idealized explanation}, which compares the temperatures and densities of the source clouds to temperatures and densities typical for the CGM in a cosmological simulation.
This particular data is part of the FIRE-2 public data release~\citep{wetzel2022Public} and was chosen for convenience.

On the other hand, the observers modeled \texttt{sample0} with assumptions informed by current knowledge of the CGM.
For example, the observers tested if the column densities could be explained by cooling-heating equilibrium driven by photo-ionization equilibrium, in which case the estimated temperature is calculated from the estimated metallicity and density parameters.
Under these assumptions the observers found reasonable fits to the provided column densities for sightlines 06 through 10.
Observers found reasonable fits for three of the other sightlines after fixing the density to an assumed $n_{\rm H} = 10^{-3.9}$ cm$^{-3}$ and assuming collisional ionization.
In this case, the temperature can be determined, but the metallicity also depends on the density,
which was not actually close to the assumed value.

In the revised modeling observers removed any assumptions of cooling-heating equilibrium, and found much better agreement with the source properties.
Even still, in four of the ten sightlines (\textsc{01}, \textsc{02}, \textsc{06}, \textsc{07}) the best-fit parameters resulted in failed CLOUDY runs with the error being ``optically thick to electron scattering - Cloudy is not intended for this regime'', which required observers to identify and scale solutions with the same ion densities but less total mass.

To summarize, the \texttt{sample0} models were not realistic based on the parts of parameter space occupied by the synthetic data and the initial parameter estimation assumptions were too rigid.
The lesson learned is that there will be places in the real universe where heating and cooling do not balance, and the solution that observers obtain may not be unique.
This is particularly true for higher ionization gas (at $T \gtrsim 10^{5}$ K and/or $n_{\rm H} \lesssim 10^{-2}$ cm$^{-3}$).
Note that the effects shown here may be stronger because only ion column densities were provided (as opposed to full spectra).

\subsection{Spectra of multi-cloud systems --- \texttt{sample1}}
\label{s: results -- sample1}

In Figure~\ref{f: sample1 spectrum} we show an example spectrum provided to observers as part of \texttt{sample1}, and the VP from the parameter-estimation best fits.
The grey bands indicate areas that are masked and do not contribute to parameter estimation,
either because there is a blend known to contaminate the spectrum or because there is no detected absorption in the region.
For \texttt{sample1} the parameter-estimation best fit was selected from models with varying numbers of clouds.
In the displayed sightline, the data is best fit by two clouds, one with significant low-ion absorption (e.g. \ion{C}{III})  and one with significant high-ion absorption (e.g. \ion{O}{VI}).
In \texttt{sample1} the observers did not perform a revised parameter estimation because the blinded parameter estimates agreed well with the source data, in part because cooling-heating equilibrium was not assumed.

Figure~\ref{f: sample1 violin} compares the posteriors from the parameter estimation to the properties of the source clouds.
Posterior probability distributions, or posteriors for short, describe the predicted probability of the source properties having a given value.
The posteriors are displayed via a violin plot,
where the width of the vertical distribution at a given value scales with the posterior probability distribution at that value.
Violin plots are equivalent to histograms rotated 90$^\circ$ and reflected to be symmetric.
In most cases the source properties reside well inside the posteriors, i.e. the posteriors are typically sufficiently extended.
As with the blind modeling for \texttt{sample0} some actual values lie outside the assumed range of possible values: 
some source clouds have $Z > 10^{1.5} Z_\odot$, while the parameters were estimated assuming $Z$ spans $\log Z/Z_\odot = [-3, 1.5]$ (where $\log Z/Z_\odot = 1.5$ is the default maximum metallicity for Cloudy).
However, the source $Z$ are within $< 0.5$ dex of the highest estimated metallicity, significantly less discrepant than blind modeling for \texttt{sample0}.

The most-extensive posteriors in \texttt{sample1} are often associated with a hotter phase tracing the OVI for a few reasons.
First, the cooling of such a gas phase resembles that of collisional time-dependent gas cooling with no external radiation, and hence is independent of density.
Second, following the lessons learnt from \texttt{sample0}, the observers relaxed the priors such that the lowest temperature could be $10^2$ K, while the source data has a lower bound of $10^4$ K.
The most poorly-constrained priors extended beyond the lower bound of the source data.
Additional potential contributing factors to the wide posteriors include the absence of \ion{C}{IV} lines in the provided spectra (\ion{C}{IV} 1548, 1550 probe intermediate temperatures and are not covered by COS G130M or G160M at $z=0.25$).

The modeling identifies the correct number of components in 9 of 10 sightlines.
The exception is sightline 050, which has a hot, low-density, low-metallicity absorber that produces little absorption in the ions provided to the observers.

\subsection{Spectra of high-resolution turbulent mixing --- \texttt{sample2}}
\label{s: results -- sample2}

Figure~\ref{f: sample2 ray 15} shows an example sightline (sightline 0015) passing through the simulation used to generate \texttt{sample2}.
Figure~\ref{f: sample2 spectrum 15} shows the resulting mock spectra,
the best fit from parameter estimation,
and the posteriors for derived properties.
The best fit for sightline 0015 uses five clouds.
Figure~\ref{f: sample2 15} compares the parameter-estimation posteriors for the sightline to the source property distributions.
Both the source data and the posteriors are weighted by associated \ion{H}{I} column density.
Integrating over the source data for any property yields the total \ion{H}{I} column density.
Integrating over any property posterior for a component yields the median estimated \ion{H}{I} column for that component.
The combined posterior is the result of co-adding the normalized single-component distributions.
As mentioned previously, the properties of cool gas are easier to estimate with UV absorption spectra than the properties of hot gas,
so weighting by \ion{H}{I} focuses the comparison on the regime parameter estimation can probe best.
Future work could focus on recovery of warm or hot phases by weighting by other ions.

In our analysis we typically bin the posterior distribution and use the peak as the maximum likelihood estimate (MLE).
An alternative is to use a ``point-estimate'' MLE, i.e. the single set of parameters with the highest probability across all sets tested during parameter estimation.
We do not use the point-estimate MLE because it can be noisy and not representative of the posterior, but Figure~\ref{f: sample2 spectrum 15} shows this estimate as stars.

In \texttt{sample2} the ``original blinded'' data provided to the observers was a $z=2$ simulation ``observed'' at $z=0.13$ (\S\ref{s: data generation -- sample2}).
This choice complicated the interpretation of the parameter estimates,
so the observers reran their parameter estimation on the same sightlines but now ``observed'' at $z=2$.
In our analysis we utilize the ``rerun'' parameter estimates,
and relegate the original blinded parameter estimates to Appendix~\ref{a: sample2 blinded}.
We emphasize that, while not strictly blinded, the observers did not revise their methodology prior to analyzing \texttt{sample2} ``observed'' at $z=2$.

The MLE of the combined distribution is usually dominated by the strongest component,
and agrees relatively well with the peak values of the source data.
The lower-column-density components are poorly constrained, but do enclose the source data.

Figure~\ref{f: sample2 structure 15} shows for sightline 0015 the relationship between gas properties as a function of line-of-sight position vs as a function of line-of-sight velocity.
The gas sightline 0015 passes through in Figure~\ref{f: sample2 ray 15} shows up as $\approx 5$ detectable regions in $v_{\rm LOS}$-$x$ space (the top left panel of Figure~\ref{f: sample2 structure 15}).
This is the same number of regions as there are components derived from parameter estimation,
though there is not a clear correlation between the regions and the number of components.
Two of these regions are responsible for the majority of the \ion{H}{I} absorption,
and the parameter estimation describes these regions with a $\log N_{\ion{H}{I}} \approx 17.2$ component (yellow) and a $\log N_{\ion{H}{I}} \approx 17$ component (pink).
The density, temperature, and metallicity of these two strongest components are well-constrained, and consistent with the source data.
The parameter estimation also correctly predicts these components as having kpc-scale pathlengths (Figure~\ref{f: sample2 spectrum 15} lower right).
The properties of the remaining, weaker absorption, components are not as well constrained but conservative enough to contain the source values ($n_{\rm H}$, $Z$, $T$) within $\sim 1$ dex.
Note that while the absorbing gas spans $-7 {\rm kpc} \lesssim x \lesssim 15$ kpc the ($n_{\rm H}$, $Z$, $T$) distributions are similar throughout.
To rephrase, cloud complexes have similar ($n_{\rm H}$, $Z$, $T$) distributions---there are not ``cool'' clouds with distinct properties.

Figure~\ref{f: sample2 violin} extends the comparison between estimated parameters and source data properties from one sightline to the full sample.
Each violin shape is a 1D distribution like those shown in Figure~\ref{f: sample2 15}, rotated 90 degrees.
The linear scale used here emphasizes the dominant $v_{\rm LOS}$, $Z$, $T$ and $n_{\rm H}$ for both the source and parameter estimation.
The shape of the $v_{\rm LOS}$ posteriors closely traces the actual distribution of $v_{\rm LOS}$ for both parameter estimations.

Figure~\ref{f: sample2 violin vs components} reveals the extent to which parameter estimation can capture the structure beyond the dominant components.
Each violin is proportional to the \textit{logscale} 1D source-data distribution,
where only bins that contain $N_{\ion{H}{I}} > 10^{12.5}$ cm$^{-2}$ are shown.
This value was selected because there are few parameter-estimation components with lower column densities.
For example, for $\log Z$ we set the thickness of the violin to zero for $\frac{dN_{\ion{H}{I}}}{d\log Z} < 10^{12.5} {\rm cm}^{-2} / \Delta \log Z \approx 10^{12.5} {\rm cm}^{-2} / 0.145 \approx 2.2 \times 10^{13}$ cm$^{-2}$.
The points show the MLEs for each component the parameter-estimation fit is composed of,
with the opacity scaled logarithmically by $N_{\ion{H}{I}}$ of the component.
The components become completely invisible as $N_{\ion{H}{I}} \rightarrow 10^{12.5}$ cm$^{-2}$.
The MLEs of the components with the highest $N_{\ion{H}{I}}$ contributions agree well with the properties at the peaks of the source data.
However, MLEs of some components describe gas that is $\sim 1$ dex cooler, denser, and more metal-poor than present in the source data.
Note that the posterior percentiles of such components still overlap the source data.
In the simulations this gas does not exist because the UVB provides an effective temperature floor,
below which CGM gas has trouble cooling because it is photo-heated by background radiation.
However, following \texttt{sample0} the decision was made to perform parameter estimation without restricting the parameter space by assuming cooling-heating equilibrium.
This is discussed further in \S\ref{s: discussion -- cloud structure}.

\section{Discussion}
\label{s: discussion}

\subsection{The cloud structure of absorbing gas}
\label{s: discussion -- cloud structure}

\subsubsection{Extent of recovery}
\label{s: discussion -- cloud structure -- recovery}

The goal of parameter estimation is not just to estimate the average properties of absorbing gas,
but also to disentangle the contributions of multiple clouds.
The modeling challenge spans three regimes of ``cloud structure'':
a single uniform cloud,
multiple uniform clouds,
and multiple distributions of clouds.
Many quantities relevant to constraining the physics of the CGM only require cloud structure to be separated on the level of one or multiple uniform clouds.
For example, metallicity can help determine if absorbing gas is pristine inflow or enriched outflow without requiring resolving many clouds~\citep[e.g.][]{hafen2017Lowredshift}.
On the other hand, the role complex multiphase gas plays in the physics of the CGM has been a focus of much recent research~\citep[e.g.][]{voit2015Precipitationregulated, esmerian2021Thermal, smith2023Arkenstone, tan2023Cloudy},
and constraining this observationally may require identifying cloud structure on the level of multiple distributions of clouds.

Our analysis confirms that parameter estimation can successfully extract the properties of uniform clouds.
This is apparent from the revised \texttt{sample0} estimates.
Parameter estimation was similarly successful at disentangling multiple uniform clouds (\texttt{sample1}).

For \texttt{sample2} with its multiple distributions of clouds we ask whether parameter estimation can recover:
a) the primary velocity components,
b) the primary metallicities,
c) the primary ionization states, and
d) the shapes of the property distributions.
Regarding capturing the bulk of the velocity structure, Figure~\ref{f: sample2 violin vs components} shows that in all \texttt{sample2} sightlines the component best estimates span the full width of the $v_{\rm LOS}$ distribution.
Note that the error bars show only the 68th percentile of the component posteriors (see Figure~\ref{f: sample2 15} for an example of the range of the full posteriors).

Concerning metallicity, in all \texttt{sample2} sightlines the metals were well-mixed, resulting in a narrow metallicity distribution peaked at {\metallicity} $\sim -1.2$.
Parameter estimation identified at least one component consistent with this metallicity in all sightlines.
However, in $\sim 6$ / 10 sightlines parameter estimation also identified a significantly-contributing component with a metallicity $>0.5$ dex higher or lower than the source metallicity,
albeit with 16th-84th percentile uncertainties that are typically consistent with the source metallicities.

Regarding the main ionization states of distributions of clouds,
the parameter estimation captures the majority of the low-ionization gas well in all three samples, but the properties of $T \gtrsim 10^{4.5}$ K gas are more poorly-constrained.
In \texttt{sample1} the components with the largest uncertainties were $T \gtrsim 10^5$ K gas,
and the only cloud not identified had $T \sim 10^{5.7}$ K.
In \texttt{sample2}, all sightlines identify the $T \sim 10^4$ K gas that dominates the absorption (Figure~\ref{f: sample2 violin vs components}),
and find that additional hard-to-constrain components improve the fit.
Roughly half the sightlines suggest low-absorption $T \sim 10^3$ K gas may be present, albeit with uncertainties $\gtrsim 0.5$ dex, which is a temperature not present in the simulation due to heating from the UVB.
Regarding the hotter gas, the hot gas parameters are especially difficult to estimate in \texttt{sample2} because the pathlength necessary to accrue significant absorption is longer than any sightline through the \texttt{sample2} simulations.
Difficulty recovering higher-temperature gas can also result from lack of access to high-ionization absorption lines such as \ion{Ne}{VIII} and \ion{Mg}{X},
although in \texttt{sample2} the observers had access to these lines.
In addition, at $T \gtrsim 10^5$ K, gas cooling becomes independent of radiation field, and hence the density. 
Therefore, the inferred properties for high temperature gas are generally more uncertain.

Finally, regarding the shapes of the property distributions,
the employed parameter estimation approximates the absorption as arising in several uniform clouds,
and as such fundamentally cannot constrain distribution shape.
This is an area of future work that some alternative parameter-estimation methods address~\citep[such as fitting the underlying density distribution as a power law;][]{stern2016Universal}.
Another path to explore is to model each absorption system as a linear combination of turbulent-mixing-layer absorption profiles~\citep{tan2021Model}, $T \sim 10^4$ K absorption profiles, and $T \sim 10^6$ K absorption profiles. 

\subsubsection{Factors affecting recovery}
\label{s: discussion -- cloud structure -- factors}

The velocity-space properties of the source gas significantly affected parameter estimation.
All sightlines in \texttt{sample2} had very similar source $Z$, $T$, and $n_{\rm H}$ distributions,
similar column densities ($N_{\ion{H}{I}} \sim 10^{17}-10^{18}$ cm$^{-2}$),
and widely varying source $v_{\rm LOS}$ distributions. 
As such, \texttt{sample2} provides a test of how line-of-sight velocities affect parameter estimation.
From Figure~\ref{f: sample2 violin vs components} we can see that varying $v_{\rm LOS}$ distributions drive significant variety in recovery of $Z$, $T$, and $n_{\rm H}$, particularly for weak, low density, high temperature components.

The ions available for fitting determine much of how well cloud structure can be retrieved.
Utilizing multiple Lyman-series lines reduces the contamination from noise during identification of coincident-in-velocity-absorption.
Intermediate ionization states, especially strong ions such as \ion{C}{IV}, play an important role in separating gas phases because they often have contributions from both a lower and a higher ionization transition.
This is shown with the \ion{C}{III} absorption in Figure~\ref{f: sample1 spectrum}.

The \texttt{sample2} analysis probes the regime of fractal cloud structure,
which provided a challenge for parameter estimation built on uniform clouds.
The gas probed in \texttt{sample2} consists of both $T \sim 10^4$ K and $T \sim 10^6$ K gas,
and turbulent mixing layers at the interface of the cool and hot gas.
Analytic theory and numerical simulations predict that each turbulent mixing layer has a fractal structure~\citep[e.g.][]{fielding2020Multiphase, Kanjilal.etal.2021, tan2021Model, Gronke.etal.2022}.
With a fractal structure any sightline that pierces a turbulent mixing layer will probe a distribution of temperatures and densities, regardless of the velocity resolution of the spectra.

The challenges found here are present despite performing parameter estimation under optimal conditions:
for \texttt{sample2} the lines available to the observers provided extensive coverage of multiphase gas, the provided data had a high signal-to-noise ratio of 30, and absorption was not contaminated by absorption associated with cosmological structures other than the source.
However, some scenarios may enable better recovery of properties than the scenarios tested here, e.g. if the many small clouds have narrower profiles or are more widely distributed in velocity space than in \texttt{sample2}.
Related, in \texttt{sample0} the theorists provided ion column densities instead of full spectra.
While this simplified the data it also removed important information.

\subsubsection{Implications for existing results}
\label{s: discussion -- cloud structure -- implications}

In several studies, it has been found that small, dense clouds are required to explain observed absorption profiles~\citep{Rigby2002, Narayanan2008, Muzahid2018}.
Across all samples in the present study the density of the strongest absorbing component is not overestimated,
suggesting that the high density estimates are not a common artifact of parameter estimation.
Some of the components nonessential for a good fit can have overestimated best-estimates for density---in 4 / 10 of the sightlines of \texttt{sample2} there exists such a component where the best estimate for the density is $\sim 1-2$ dex higher than found in the source data.
This suggests that the best estimate of poorly-constrained components is not good evidence for small, dense gas clouds.
However, the 16th-84th percentile uncertainties of these components overlap the source density,
so these components are not discrepant with the source data.
We reiterate that none of the strong, well-constrained components have over-estimated densities.

Grid-based simulations such as those employed here can struggle to simulate high-density gas.
This raises the question of whether the immediately-above conclusion regarding lack of density overestimates is subject to numerical issues.
This is not the case in our analysis---the sharp cutoffs seen in the \texttt{sample2} density distributions (e.g. Figure~\ref{f: sample2 violin vs components})  at maximum densities of $n_{\rm H} \sim 10^{-2}$ cm$^{-3}$ are well below any numerical density ``ceiling'' of these simulations.
Instead these cutoffs are driven by an effective density ceiling from photoionization and pressure equilibrium.
More specifically, since the simulations included an ionizing UV background with no correction for self-shielding, gas cannot cool far below $10^4$~K, even though the cooling function does allow for cooling down to $\sim 10$~K.
Since the simulation began in pressure equilibrium and there are no strong shocks generated during the simulation (the turbulence that develops in the mixing layer is subsonic), this implies that the density cannot increase much above $n_{\rm H}\sim 10^{-2}~{\rm cm^{-3}}$ \citep[see][]{mandelker2020Instability}.

In our analysis the parameter estimation excels at recovering the properties of the dominant absorbing gas
but the properties of weaker components are poorly constrained,
possibly due to limitations in the provided spectra.
This has implications for the required resolution for generating mock observations.
Because thermally unstable gas may have a fractal structure, fully resolving the gas may be effectively impossible.
However, if spectra do not provide enough data to distinguish structure beyond the dominant properties then the resolution required to produce accurate mock observations may be less stringent.
One suggestion for the target resolution for mock observations is the resolution at which the mass of cool gas converges~\citep[e.g.][though such a scale may not exist]{mccourt2018Characteristic},
which is still well below the resolution of most cosmological simulations.
Further resolving the simulated gas may yield more complex cloud structure,
but it may also be below the resolution of the observations, particularly for very cold gas~\citep[e.g.][]{jones2010Bare}.

Some works argue that it is possible to extract detailed information about the origin of absorbing gas with the aid of additional observations of the galaxy associated with the absorption system.
For example, \cite{peroux2013SINFONI} argue that one of their identified absorption systems is dominated by gas originating in an outflow, and \cite{peroux2017Nature} suggest that a separate absorption system has little contribution from outflows.
Making statements regarding the origin or fate of absorbing gas may require being able to identify and separate multiple clouds along a given line of sight,
because sightlines are predicted to frequently penetrate multiple clouds with multiple origins~\citep[e.g.][]{hafen2019Origins, hafen2020Fates}.
Our results indicate that clouds with distinct, uniform properties can indeed be distinguished from one another successfully (Figure~\ref{f: sample1 violin}).
On the other hand, \texttt{sample2} is spectra of a filament falling in from the IGM and passing through a metal-enriched halo.
In this case individual clouds that produce weak absorption cannot be easily identified in the provided spectra,
but the average properties of the filament can be identified clearly.
and these properties may allow observers to constrain the \ion{H}{I}-weighted origin of absorbing gas if constraining origin via e.g. metallicity.

\subsection{Choices in parameter estimation}
\label{s: discussion -- modeling choices}

\subsubsection{Discrete uniform clouds versus distributions of clouds}

Modeling absorption as arising in discrete components is in part motivated by observed profiles that are often quite symmetrical,
suggesting a single region with a given density and temperature. 
This is especially relevant for colder clouds traced with low ionization absorption.
Some components are also observed to be quite narrow~\citep[e.g.][]{churchill1999Population, churchill2001Kinematics}, 
which requires cloud distributions to either be confined to specific regions of the CGM,
or for separate regions to coincidentally have similar $v_{\rm LOS}$.
The former is plausible:
idealized simulations, analytic theory, and cosmological simulations indicate that mixing and perturbations from cold gas may be primary drivers of further cold gas growth~\citep[e.g.][]{nelson2020Resolving, esmerian2021Thermal, Gronke.etal.2022, gronke2022Cooling, ramesh2022Circumgalactic, saeedzadeh2023Cool}.
This suggests that cool cloud distributions may indeed be limited to regions of the CGM with adjacent or already-present cool gas.

The most realistic among our samples, \texttt{sample2}, consists not of discrete clouds, but of a complex cloud structure with a distribution of properties.
This is consistent with analytic theory and high-resolution simulations that predict turbulent mixing layers have fractal structure with a continuous distribution of temperatures~\citep[e.g.][]{tan2021Model}.
The implication is that any fully-successful model of absorption systems should also include absorption from gas with a distribution of properties.
Such models exist~\citep[e.g.][]{stern2016Universal}, but are not commonly employed in parameter estimation.
Given a physically-motivated parameter distribution (e.g. the temperature distribution expected in a mixing layer), these models have the potential to significantly improve the accuracy of the reproduced parameters.

\subsubsection{Assumptions about thermal and ionization equilibrium}

At the first step of our analysis the observers assumed cooling-heating equilibrium.
This assumption resulted in inaccurate estimates of the parameters of \texttt{sample0}, though the magnitude of the errors was significantly exacerbated by the theorists providing source properties with unrealistic temperatures and densities.
Removing this assumption produced very good fits to the data regardless of temperature-density phase-space location.
In \texttt{sample2}, gas in the source simulation is allowed to cool below $T \sim 10^4$ K,
but photo-heating from the UVB provides an effective temperature floor.
The estimated components with the strongest absorption have estimated $T \sim 10^4$ K,
consistent with the source data despite not assuming cooling-heating equilibrium.
This suggests that, for modeling the majority of absorbing gas, the safest option is to forgo assumptions of cooling-heating equilibrium.
However, in approximately half the \texttt{sample2} sightlines the best estimates for some individual, lower-$N_{\ion{H}{I}}$ components fall below the effective temperature floor (Figure~\ref{f: sample2 violin vs components}),
inconsistent with the source data.
This may suggest that estimating the parameters of poorly-constrained clouds without requiring the clouds to exist in a physically-expected region of parameter space can result in unlikely-to-be-real estimated values.

Throughout the analysis, the participating observers and theorists assumed ionization equilibrium, with contributions from both collisional- and photo-ionization.
On the data-generation side, this is a common assumption in cosmological and idealized hydrodynamical simulations, largely driven by the challenge in modeling non-equilibrium chemistry~\citep[e.g.][]{richings2014Nonequilibrium}.
However, in some cases non-equilibrium chemistry can have a significant effect on the properties of absorbing gas in the CGM, including providing one explanation for the observed bimodality in \ion{O}{VI} around low-redshift galaxies~\citep{oppenheimer2016Bimodality}.
The effect of ionization equilibrium has also been studied in focused simulations of turbulent mixing layers~\citep[e.g.][]{ji2019Simulations}.
Note that the ionization and recombination timescales are different for every ion.
It is possible that at a given ($Z$, $T$, $n_{\rm H}$) some ions are ionization equilibrium while others are not.
The effect of assuming ionization equilibrium on parameter estimation is a promising avenue of further research.

\subsubsection{Absorption modeling ``by hand''}

There are a number of works that use a less-algorithmic, more people-driven approach to model sightlines~\citep[e.g.][]{churchill1999Multiple, charlton2000anticipating, charlton2003high, ding2003quadruple, ding2003multiphase, ding2005absorption, zonak2004absorption, masiero2005models, lynch2007physical, misawa2008supersolar, lacki2010z, jones2010Bare, muzahid2015Extreme, richter2018New, rosenwasser2018understanding, norris2021Discovery}.
The methods employed by the observers in this work (CMBM; \citealt{sameer2021Cloudbycloud, sameer2022Probing}) are based on these earlier models but with significant improvements.
CMBM selects a set of possible properties, generates spectra, and repeats the process based on agreement with the observed spectra.
This algorithmic approach covers a wider range of parameter space than feasible when modeling ``by-hand'', including quantifying the likelihood of the range of possible models.
Additional beneficial features include
a $\sim 10-100 \times$ automation-enabled decrease in human hours,
the ability to vary all parameters independently (including temperature),
the output includes the full posterior for each cloud,
and the self-consistent inclusion of all transitions, weighted by noise level.
CMBM also handles blending well, an issue in real observations that is largely not present here
(the only line with slight blending was \ion{Si}{II} 989 from \ion{N}{III} 989, and this was included in the modeling).

\subsection{Comparison to similar analyses}

Absorbing gas with the same LOS velocity is not necessarily co-spatial, but instead a single absorption component can correspond to multiple clouds of gas that are physically separated, or a continuous distribution which spans a range of gas densities/temperatures. 
In an analysis of mock absorption spectra from cosmological simulations, \cite{marra2022Examining} show that only $\sim 30-50\%$ of absorption components are composed of a single, contiguous cloud, and $\sim 50\%$ of multiple clouds contributing to a single component are separated by $\gtrsim 3-12$ kpc depending on the ion.
Further, \cite{marra2022Examining} find that the amount of mass in common between velocity-aligned absorption from different ions varies widely, with $\sim 1/3$ of \ion{Si}{II} and \ion{C}{IV} absorbers sharing $>50\%$ of their mass.

Regarding estimates of the bulk properties of gas along a LOS,
\cite{liang2018Observing} determined that a Bayesian method of modeling absorption lines was able to derive LOS \ion{H}{I} mass-weighted average properties.
\cite{marra2021.cosmo.sims.test.observational.modeling} confirmed this finding using a blind study of spectra from cosmological simulations of a $z=1$ Milky Way-type galaxy and a $z=0$ dwarf galaxy.
Approaching this problem from the observer side yields similar results:
\cite{sameer2021Cloudbycloud} showed that the HI-weighted average metallicities of multiple components agree with the single values that earlier modeling had found by using total column densities as constraints.
However, a single valued property washes out the different physical processes that could be responsible for those different components.
A number of analyses test the robustness of density and metallicity to changes in the UV background~\citep[][]{Wotta2016, Wotta2019, acharya2021How,Gibson2022}, and find that these factors are uncertain by factors of a few.

\subsection{Recommendations}

Based on our analysis we provide brief recommendations for observers estimating the parameters of real and mock spectra and for theorists producing mock spectra.
Please note that many observers and theorists may already be following similar guidelines, 
independent of this analysis.

\subsubsection{For observers}

\begin{enumerate}
    \item Avoid restricting the parameter space of derived properties, including temperature, density, and metallicity (\S\ref{s: results -- sample0}). 
    \item The distribution of possible values is more informative than a single best value. Even when the best estimated value and the real value are not the same, the real value is typically among the likely values (Figure~\ref{f: sample1 violin}).
    \item Be cautious when identifying multiple contributing clouds---while the absorption-weighted averages can reliably be recovered, successful parameter estimation of multiple clouds depends on how distinct the cloud properties are and the per-cloud contribution to observable ion absorption (Figure~\ref{f: sample2 violin vs components}).
    \item Explore modeling absorption systems using physically-motivated non-uniform systems (e.g. turbulent mixing layers) in addition to uniform clouds (\S\ref{s: discussion -- modeling choices}).
\end{enumerate}

\subsubsection{For theorists}

\begin{enumerate}
    \item Avoid source properties that are overly pathological. While testing the limits of  parameter estimation is useful, unusual source properties are disproportionately likely to be poorly modeled (Figure~\ref{f: idealized explanation}).
    \item Provide mock spectra when available, not just column densities (\S\ref{s: results -- sample1}).
    \item No single metric is likely to adequately describe the relationship between the source and estimated properties. Using multiple visualizations of the same data can help (e.g. Figures~\ref{f: sample2 ray 15} to~\ref{f: sample2 violin vs components}).
\end{enumerate}

\section{Conclusions}
\label{s: conclusions}

In \textit{the Halo21 absorption modeling challenge} we generated ``observed'' column densities or spectra for three datasets,
and tested the ability of parameter estimation to derive the source properties ($Z$, $T$, $n_{\rm H}$, and $v_{\rm LOS}$) of the underlying gas.
We started with simple source data and ramped up the complexity over the course of the challenge.
For \texttt{sample0} theorists provided the column densities of uniform clouds, with the properties randomly selected from ($Z$, $T$, $n_{\rm H}$) 
typical values of the CGM, though the combined properties were not necessarily typical.
For \texttt{sample1} theorists provided absorption spectra of multiple uniform clouds, with properties drawn from gas distributions found in the CGM of the EAGLE cosmological simulations~\citep[e.g.][]{schaye2015EAGLE}.
For \texttt{sample2} theorists provided absorption spectra of a high-resolution simulation of a $T \sim 10^4$ K filament embedded in a $T \sim 10^6$ K halo~\citep{mandelker2020Instability}.
The sample2 spectra pierced turbulent-mixing zones, and subsequently had a complex cloud structure.
For all three samples the observers estimated the gas parameters via cloud-by-cloud, multiphase, Bayesian ionization modeling (CMBM) methods~\citep{sameer2021Cloudbycloud, sameer2022Probing}.
CMBM enabled observers to systematically explore the parameter space to identify the number of clouds and their properties that provided an acceptable fit to the data.
Our conclusions are as follows.

\begin{enumerate}
    \item \textbf{Best estimates:} Across all three samples the maximum-likelihood estimates from parameter estimation agree well with the bulk properties of the absorbing gas (Figure~\ref{f: summary}), provided temperature and density are allowed to vary independently as part of the fitting process (\S\ref{s: results -- sample0}).
    \item \textbf{Uncertainty estimation:} The inferred range of feasible gas properties is broad for weakly-absorbing components. The source parameters typically lie within the 16th to 84th percentiles of the posterior distribution (also Figure~\ref{f: summary}), suggesting errors are typically sufficiently conservative.
    \item \textbf{Recovery of detailed cloud structure:} Parameter estimation successfully recovered the number and properties of multiple uniform clouds (Figure~\ref{f: sample1 violin}). Parameter estimation captured the one or two dominant absorption components in the turbulent-mixing scenario, however the parameters of additional absorbing gas were poorly constrained (Figure~\ref{f: sample2 violin vs components}). Including multiple additional clouds beyond the dominant ones nevertheless did improve the fit, consistent with the true existence of complex cloud structures in the source. 
    \item \textbf{Highlighted factors affecting parameter estimation:} The imprint of the source line-of-sight velocity distribution on absorption spectra can strongly affect recovery of $Z$, $T$, and $n_{\rm H}$ (\S\ref{s: discussion -- cloud structure -- factors}). Ionization state also plays an important role, with better recovery of lower-ionization, $T \sim 10^4$ K gas.
\end{enumerate}

There are a number of avenues for future work.
As briefly discussed in \S\ref{s: discussion -- cloud structure -- implications},
provided the average temperature, density, and metallicity distributions are well-converged,
hydrodynamic simulations may only need to surpass the ``resolution'' of the observations and parameter estimations to produce sufficiently realistic mock spectra.
To rephrase, \textit{if} the simulations are converged with respect to bulk properties it may not be necessary to further resolve the gas structure, as spectra may not provide enough information to extract that structure.
However this warrants further analysis.

Reasonable agreement in ion column densities does not necessarily signal that the modeled properties are consistent with the actual properties (\S\ref{s: results -- sample0}, Appendix \ref{a: error vs error}).
However, this result is found under unusual circumstances---observers were only provided with ion column densities rather than spectra, and the source data lay in a region of parameter space not typically occupied by CGM gas.
A promising future direction may be to evaluate spectra goodness-of-fit as a predictor of accuracy and uncertainty in $Z$, $T$, and $n_{\rm H}$.
Such an analysis would be complementary to repeating the mock data challenge on additional data sources with properties differing from common parameter estimation assumptions,
e.g. sources such as cosmological simulations or a density power law.

Finally, as discussed in \S\ref{s: discussion -- modeling choices}, models composed of multiple \textit{non-uniform} clouds have the potential to play an important role in future parameter estimation.

\section*{Acknowledgements}

This work exists thanks to the Halo21 KITP virtual conference, organized by Cameron Hummels, Ben Oppenheimer, Mark Voit, and Jess Werk.
ZH was supported by a Gary A. McCue postdoctoral fellowship at UC Irvine.
We thank J. Xavier Prochaska for proposing the challenge.
CBH is supported by NSF grant AAG-1911233, and NASA grants HST-AR-15800, HST-AR-16633, and HST-GO-16703.
S and JC were supported by HST-AR-16607.
NM acknowledges support from ISF grant 3061/21 and BSF grant 2020302, and partial support from the Gordon and Betty Moore Foundation through Grant GBMF7392 and from the National Science Foundation under Grant No. NSF PHY1748958. 
NW was supported in part by a CIERA Postdoctoral Fellowship.
JSB was supported by NSF grant AST-1910346.
NL acknowledges support from NASA through grant HST-AR-15634 from the Space Telescope Science Institute (STScI), which is operated by the Association of Universities for Research in Astronomy, Incorporated, under NASA contract NAS5-26555.
JS was supported by the Israel Science Foundation (grant No. 2584/21) and by the German Science Foundation via DIP grant STE 1869/2-1 GE 625/17-1. 
This research was supported in part by the National Science Foundation under Grant No. NSF PHY-1748958.
Some computations for this research were performed on the Pennsylvania State University's Institute for Computational and Data Sciences' Roar supercomputer.
This work used the DiRAC@Durham facility managed by the Institute for Computational Cosmology on behalf of the STFC DiRAC HPC Facility (www.dirac.ac.uk). The equipment was funded by BEIS capital funding via STFC capital grants ST/K00042X/1, ST/P002293/1, ST/R002371/1 and ST/S002502/1, Durham University and STFC operations grant ST/R000832/1.
DiRAC is part of the National e-Infrastructure.

\section*{Data Availability}

The FIRE-2 simulation data used in Figure~\ref{f: idealized explanation} is available as part of a public data release~\citep{wetzel2022Public}.
The EAGLE simulation data \citep{EagleTeam2017} and halo catalogues \citep{McAlpine2016} are publicly available. 



\bibliographystyle{mnras}
\bibliography{hafen_references, wijers_references, newerref} 



\appendix

\section{Goodness of fit as an indicator of parameter estimation accuracy}
\label{a: error vs error}

\begin{figure*}
    \centering
    \includegraphics[width=\textwidth]{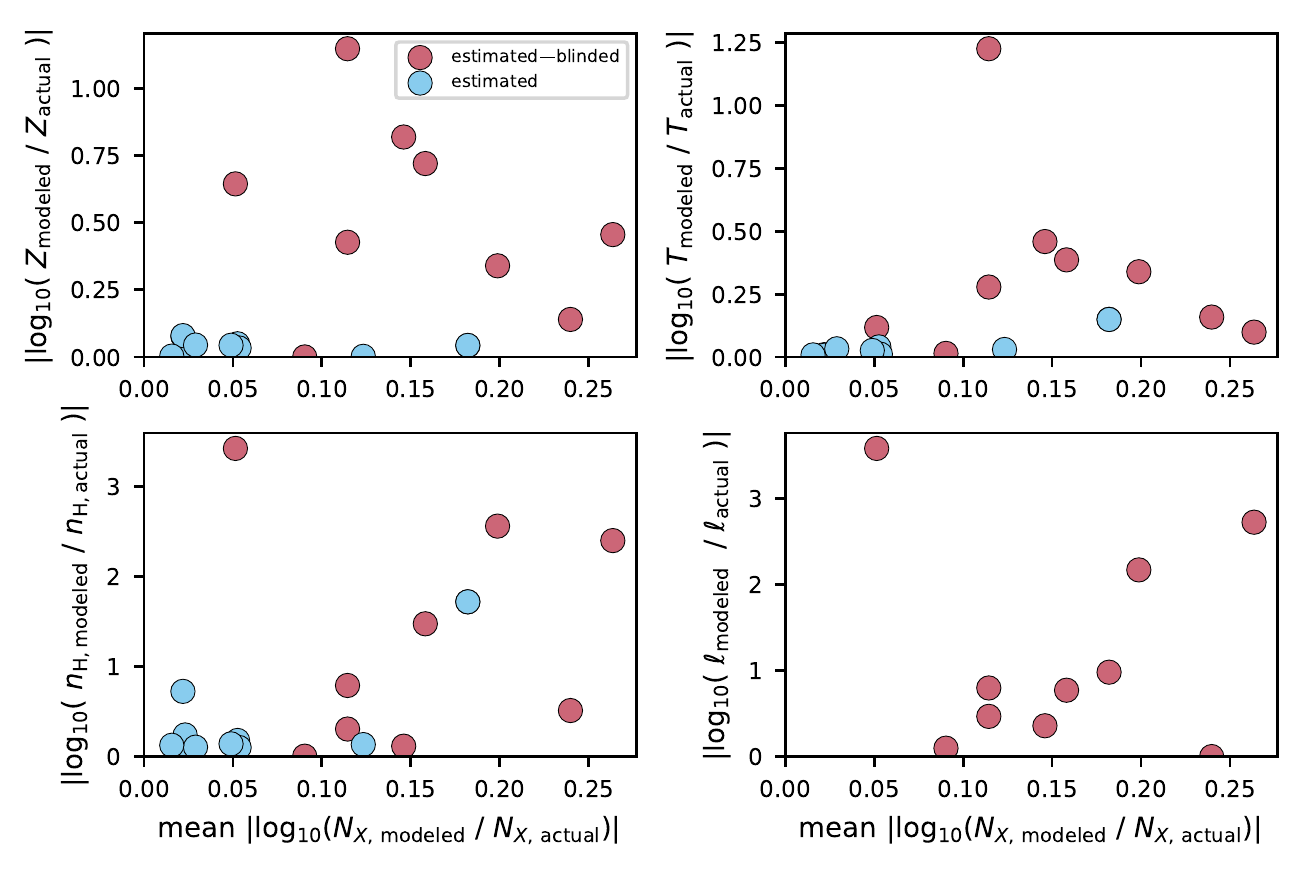}
    \caption{
    Error in properties of the absorbing gas vs error in column densities for \texttt{sample0}.
    The x-axis shows average log-space difference between column densities across all ions, weighted by the provided uncertainty.
    The original models fit the data decently (mean $\vert \log( N_{X,\,{\rm modeled}}$ / $N_{X,\,{\rm actual}}) \vert < 0.3$), but do not accurately recover the properties of the absorbing gas (e.g. $\vert \log( n_{\rm H,\,modeled} / n_{\rm H,\,actual}) \vert \gtrsim 1$).
    The revised models (which do not assume cooling-heating equilibrium) recover the properties of the gas much better (e.g. $\vert \log( Z_{\rm modeled} / Z_{\rm actual}) \vert \lesssim 0.1$), and typically fit the data to mean $\vert \log( N_{X,\,{\rm modeled}}$ / $N_{X,\,{\rm actual}}) \vert \lesssim 0.05$.
    }
    \label{f: error vs error}
\end{figure*}

Analysis of \texttt{sample0} (\S\ref{s: results -- sample0}) suggests reasonable agreement between modeled and actual ion column densities is possible, even when the modeled temperature, density, and metallicity are inconsistent with the actual values.
We demonstrate this in Figure~\ref{f: error vs error},
which compares the per-sightline disagreement between the estimated and provided properties to the disagreement between the estimated and provided column densities, averaged across all fit ions. 
The revised estimates are in much better agreement with the actual properties, but the revised ion column densities are adjusted by only $\lesssim 0.2$ dex relative to the column densities modeled blind.
More broadly, this exemplifies the issue where multiple inconsistent parameter estimates provide equally good descriptions of the data.

\section{\texttt{sample2} blinded parameter estimation}
\label{a: sample2 blinded}

\begin{figure*}
    \centering
    \includegraphics[width=\textwidth]{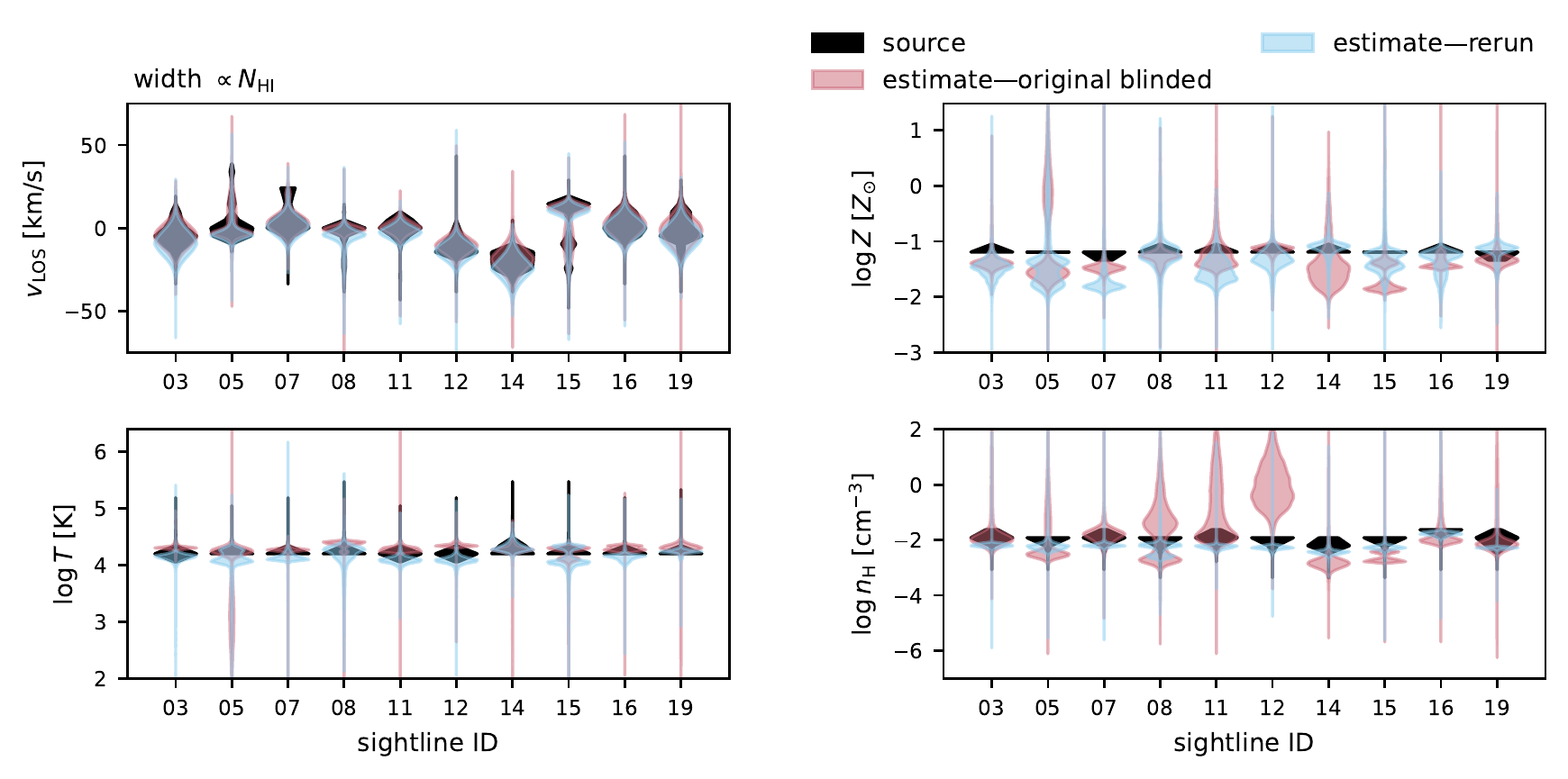}
    \caption{
    The distributions of $v_{\rm LOS}$, $Z$, $T$ and $n_{\rm H}$ for the synthetic source data (black) compared to the full posteriors from parameter estimation.
    The width of the violin plots scales with $N_{\ion{H}{I}}$.
    The posteriors for both the blinded and revised parameter estimations overlap with the source distributions in most cases, suggesting the parameter estimation is typically sufficiently conservative.
    Exceptions are
    both the blinded and revised posteriors of $Z$ for \texttt{07},
    the blinded posteriors of $Z$ for \texttt{14} and \texttt{15},
    most of the blinded posteriors of $T$,
    and both the blinded and revised posteriors of $n_{\rm H}$ for \texttt{12}, \texttt{14}, and \texttt{15}.
    }
    \label{f: sample2 violin both}
\end{figure*}

\begin{figure*}
    \centering
    \includegraphics[width=\textwidth]{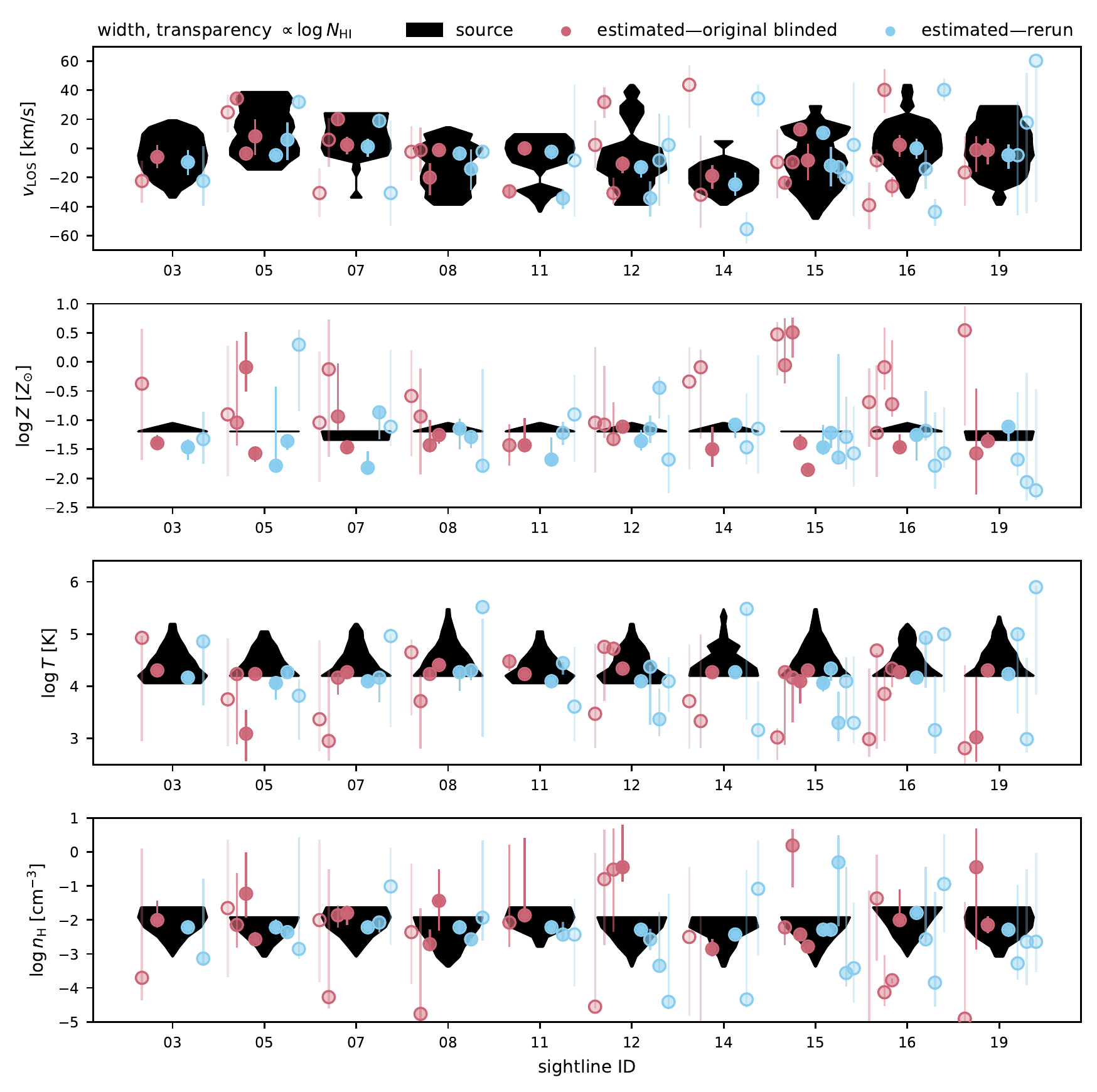}
    \caption{
    The distributions of $v_{\rm LOS}$, $Z$, $T$ and $n_{\rm H}$ for the synthetic source data (black) compared to the best estimates for individual components comprising the parameter estimation.
    Each point is the maximum-likelihood estimate (MLE) of an individual component,
    and the associated bars enclose the the 16th to 84th percentiles of the posterior.
    The width of the violin plots and the opacity of the points are scaled \textit{logarithmically} with $N_{\ion{H}{I}}$,
    such that if the parameter estimation perfectly agreed with the data all visible MLE points would lie within the source data violins.
    The MLE of the strongest component typically aligns with the peak of the source distribution, but can be off by $\lesssim 1$ dex for $Z$ and $n_{\rm H}$, especially for the blinded estimates.
    The best fit for parameter estimation consistently includes lower-column-density components with properties not found in the source data.
    }
    \label{f: sample2 violin vs components both}
\end{figure*}

For \texttt{sample2} the primary difference between the initial blinded estimates and the rerun estimates lies in the source data and the employed UVB,
rather than the estimation methodology.
The temperature, density, metallicity, LOS velocity, and total column for \texttt{sample2} were extracted from $z=2$ sightlines passing through a high-resolution simulation of a turbulent mixing zone.
To avoid providing any hints about the origin of the data the synthetic generators aimed to provide $z \sim 0 - 0.25$ spectra,
as was done for the preceding samples.
To this end, for the blinded \texttt{sample2} the synthetic generators took the $z=2$ absorbing-gas properties,
generated spectra using the $z=0.13$ UV ionization table,
and provided synthetic observations mimicking COS observations of gas at $z=0.13$.
The concept was to generate absorption from $z=0.13$ gas,
independent of the mechanisms responsible for setting the properties of the gas.
This proved to complicate the interpretation of the results, so we provided a revised set of spectra to the observers, for them to rerun their analysis on.

The revised \texttt{sample2} synthetic spectra used a $z=2$ ionization table,
had a wavelength resolution of 0.05 \AA,
a line spread function with a FWHM of 7 {\kms},
and included the ions \ion{H}{I}, \ion{Si}{II}, \ion{Si}{III}, \ion{Si}{IV}, \ion{C}{II}, \ion{C}{III}, \ion{C}{IV}, \ion{O}{I}, \ion{O}{VI}, \ion{N}{II}, \ion{N}{V}, \ion{Fe}{II}, \ion{Mg}{X}, and \ion{Ne}{VIII}.
The ($Z$, $n_{\rm H}$, $T$, $\NHI$) of the source data were not modified.
These changes followed discussion between the theorists and observers.
The better agreement between the revised parameter estimation and the source properties is likely related to the changes in the provided spectra.
For example, the blinded estimates of $T$ are consistently high by $\sim 0.1$ dex,
while the rerun estimates agree better.

\bsp	
\label{lastpage}
\end{document}